\documentclass{article}
\usepackage{graphicx} % Required for inserting images
\usepackage{amsmath, amsfonts, amssymb}
\usepackage{natbib}
\usepackage[colorlinks]{hyperref}
\usepackage[color=yellow]{todonotes}
\usepackage{tikz,tikz-cd}
\usepackage{parskip}
\usepackage[margin=2cm]{geometry}
\usepackage{subcaption}
\usepackage{float}
\usepackage{authblk}

\newtheorem{theorem}{Theorem}

\newcommand{\bjg}[1]{{\color{blue}BJG: {#1}}}
\title{Geometric derivation and structure-preserving simulation of quasi-geostrophy on the sphere}
\author[1]{Erwin Luesink}
\author[2]{Arnout D. Franken}
\author[3]{Sagy R. Ephrati}
\author[2,4]{Bernard J. Geurts}
\affil[1]{Korteweg-De Vries Institute, University of Amsterdam, PO Box 94248, Science Park 107, 1090 GE Amsterdam, The Netherlands}
\affil[2]{Multiscale Modelling and Simulation, Department of Applied Mathematics, Faculty EEMCS, University of Twente, PO Box 217, 7500 AE Enschede, The Netherlands}
\affil[3]{Department of Mathematical Sciences, Chalmers University of Technology and University of Gothenburg, 412 96 Gothenburg, Sweden}
\affil[4]{Multiscale Physics, Center for Computational Energy Research, Department of Applied Physics, Eindhoven University of Technology, Eindhoven, The Netherlands}

\date{\today}

\begin{document}

\maketitle

\begin{abstract}
We present a geometric derivation of the quasi-geostrophic equations on the sphere, starting from the rotating shallow water equations. We utilise perturbation series methods in vorticity and divergence variables. The derivation employs asymptotic analysis techniques, leading to a global quasi-geostrophic potential vorticity model on the sphere without approximation of the Coriolis parameter. The resulting model forms a closed system for the evolution of potential vorticity with a rich mathematical structure, including Lagrangian and Hamiltonian descriptions. Formulated using the Lie-Poisson bracket reveals the geometric invariants of the quasi-geostrophic model. Motivated by these geometric results, simulations of quasi-geostrophic flow on the sphere are presented based on structure-preserving Lie-Poisson time-integration. We explicitly demonstrate the preservation of Casimir invariants and show that the hyperbolic quasi-geostrophic equations can be simulated in a stable manner over long time. We show the emergence of longitudonal jets, wrapped around the circumference of the sphere in a general direction that is perpendicular to the axis of rotation.
\end{abstract}

\section{Introduction}
In geophysical fluid dynamics, one deals with fluids that have large horizontal length scales and comparably small vertical length scales. The three-dimensional Navier-Stokes or Euler equations for an incompressible, stratified fluid on a rotating domain with gravity are basic models that represent a complete representation of the dynamics. However, for practical flow on a sphere of geophysical relevance, these equations are too expensive to solve computationally for such shallow domains. One of the common approximations to reduce this computational cost is to disregard the vertical velocity entirely, which leads to the primitive equations. The primitive equations are a three-dimensional model that is still computationally costly to fully resolve. A further approximation is to consider vertically averaged equations. This procedure leads to the rotating shallow water equations, which are a compressible two-dimensional model. This approximation procedure can be found in detail in \cite{holm2021stochasticb}. The compressibility of the rotating shallow water model indicates the presence of waves, which in this particular context are called gravity waves (rather than acoustic waves in the usual compressible fluids setting). Gravity waves pose a severe restriction on the time step, since the typical velocities of the fluid flow are two orders of magnitude smaller than the velocities associated with these waves. The gravity waves can be filtered out by expanding the rotating shallow water equations around the geostrophic balance. This leads to the quasi-geostrophic equations that we focus on in this work. 

The quasi-geostrophic equations form a cornerstone in geophysical fluid dynamics at planetary scale. The quasi-geostrophic equations are typically described in a planar geometry with approximations to the Coriolis parameter, such as the $f$ and the $\beta$-plane approximation as shown in \cite{pedlosky2013geophysical, vallis2017atmospheric, zeitlin2018geophysical}. While these approximations have served well for mid-latitude atmospheric and oceanic dynamics, they do not form a global model. A global version of the quasi-geostrophic equations on a spherical domain can be derived to avoid approximations to the Coriolis parameter. This global quasi-geostrophic model is also known as the global barotropic vorticity equations and was derived earlier in \cite{verkley2009balanced, schubert2009shallow} through physical reasoning, based on considerations of \cite{cressman1958barotropic, lorenz1960energy, daley1983linear}. In this paper, we employ methods of asymptotic analysis to mathematically derive the quasi-geostrophic equations on a sphere. This approach recovers the results of \cite{verkley2009balanced, schubert2009shallow} and establishes precise conditions on the validity of the model. In particular, asymptotic methods replace physical reasoning by perturbation series expansions involving the relevant dimensionless numbers. Under the condition that certain dimensionless numbers are small enough, one can derive the quasi-geostrophic equations on the sphere and simultaneously obtain additional information on the conditions for the size of the dimensionless numbers for which the derivation can be justified.

The quasi-geostrophic equations can be expressed in a variational and in a Hamiltonian framework. In \cite{holm1998hamilton}, the quasi-geostrophic equations were derived on a $\beta$-plane in Lagrangian coordinates. This was an important development since it showed that the quasi-geostrophic equations have a rich geometric content. Using the Euler-Poincar\'e formulation that was introduced in \cite{holm1998euler}, in this paper we present a derivation of the quasi-geostrophic equations on the sphere in Eulerian coordinates. The geometric framework provides several benefits. First of all, it provides the means to characterise the conservation laws that are maintained in the quasi-geostrophic model on the sphere, including energy and enstrophy. Secondly, the mathematical structure associated with the Hamiltonian description provides important numerical advantages when simulating complex jet structures that form on a rotating sphere. The key driver that enables the consideration of the full QG model on the sphere is the fact that a scalable structure-preserving numerical method for solving the two-dimensional Euler equations on the sphere has recently become available. This method was introduced in \cite{modin2020casimir} and is based on the self-consistent finite mode truncation that \cite{zeitlin2004self} developed for hydrodynamics on the (rotating) sphere. In \cite{cifani2023efficient} the method of \cite{modin2020casimir} was parallellised to apply to high-fidelity numerical studies of two-dimensional turbulence. In \cite{cifani2022casimir}, it was then shown that this method is successful in providing numerical evidence of the double cascade mechanism. This double cascade mechanism was shown theoretically to exist for the sphere in \cite{lindborg2022two}, based on the seminal work of \cite{kraichnan1967inertial}. All these developments were, up to now, dedicated to the Euler equations and the Navier-Stokes equations on the sphere. The quasi-geostrophic equations fall into the same topological class of as the Euler equations after accounting for changes in the geometry. We elucidate this in the present paper and consequently derive the QG model on the sphere, as well as indicate a corresponding structure-preserving simulation method. As a particularly challenging test case we show the induced jet dynamics corresponding to unforced flow that was set in motion initially and which is maintained indefinitely since the numerical method has no artificial dissipation whatsoever.

The organization of this paper is as follows. In section \ref{sec:derivation} we recall the derivation of the quasi-geostrophic potential vorticity in detail where we focus on asymptotic considerations. In section \ref{sec:lagrangian} we discuss the Lagrangian and Hamiltonian functionals that are associated with the quasi-geostrophic equations on the sphere. This introduces the Lie-Poisson bracket of two-dimensional hydrodynamics on the sphere. An example simulation of unforced dynamics is presented in section \ref{sec:numerics} using the numerical model introduced in \cite{franken2023zeitlin}. In section \ref{sec:conclusion} we conclude the paper.

\section{Evolution equation for the quasi-geostrophic potential vorticity}\label{sec:derivation}
In this section, we present the main theoretical result of this paper and derive the equations for quasi-geostrophic flow on a spherical domain. Throughout, we adopt the colatitude coordinate system \eqref{eq:ds2}. In particular, we show in this section that, contrary to the planar case, the sphere admits nonlinear dependence of the flow on the Coriolis parameter while still being compatible with geostrophic balance (i.e., the balance of Coriolis force with inertial forces). We derive the quasi-geostrophic equations starting from the rotating shallow water equations. We first introduce the Lagrangian that corresponds to the rotating shallow water equations.

We express the rotating shallow water Lagrangian in dimensionless form on a rotating sphere with radius $a$. The Coriolis parameter is given by $f=2\Omega\cos\theta$, where $\Omega$ denotes the frequency. In dimensionless form, the Lagrangian for rotating shallow water is given by \cite{holm1999euler}
\begin{equation}\label{eq:lagrsw}
    L_{rsw} = \int_{S^2} \left(\frac{1}{2}|u|^2 + \frac{l}{a\,{\rm Ro}}u\cdot R - \frac{1}{2\,{\rm Fr}^2}(h-2b)\right)h\,dA,
\end{equation}
where $u$ is the velocity field, $R$ is the vector potential for the Coriolis parameter, i.e. $\nabla^\perp\cdot R = f$, $h$ is the layer thickness (or depth in oceanography), $b$ is the orography (or bathymetry in oceanography). The global Rossby number ${\rm Ro}=U/{2\Omega l}$ is the ratio between the typical horizontal velocity $U$ and velocity induced by the rotation of the sphere at the equator, where $l$ is the horizontal length scale. The dimensionless factor $\frac{l}{a}$ converts the horizontal length scale in the global Rossby number into the radius of the sphere. The Froude number ${\rm Fr}= U/\sqrt{gH}$ is the ratio between the typical horizontal velocity $U$ and the speed of the fastest gravity wave $\sqrt{gH}$, with $g$ the gravitational acceleration and $H$ the typical thickness of the fluid layer. The measure $dA = a^2\sin\theta\,d\theta\,d\lambda$ is the area measure on the surface of the sphere. 

Note that the global Rossby number does not accurately represent the balance between the inertial force and the Coriolis force. Instead, one can introduce a \textit{local} Rossby number ${\rm Ro}_{loc}(\theta) = U/fl$ to accurately capture the balance between these forces, where $f$ is the local value of Coriolis parameter. The ratio between the local and global Rossby numbers ${\rm Ro}_{loc}/{\rm Ro} = \cos \theta$ indicates that the global Rossby number accurately describes local flow conditions when $\cos \theta \approx 1$. In global geophysical fluid dynamics, the velocity scale induced by the rotation of a planet is often much larger than the horizontal velocity, leading to a small global Rossby number ${\rm Ro}\ll 1$. This implies that the local Rossby number ${\rm Ro}_{loc}(\theta)$ is also small everywhere, with the exception of a narrow band around the equator for which $\cos \theta < {\rm Ro}$. This highlights the fact that the Coriolis force vanishes at the equator.

\subsection{Euler-Poincar\'e theorem}
The Euler-Poincar\'e theorem, first introduced in \cite{holm1998euler}, is one of the cornerstones of geometric mechanics. It allows for a systematic derivations of equations in classical mechanics from energetic interpretations that involve symmetries. The Euler-Poincar\'e framework is closely related to the Euler-Arnold framework, which is named after the seminal work \cite{arnold1966principe}, in which one interprets fluid equations as geodesics on infinite-dimensional manifolds. The key difference is that equations that are derived from the Euler-Poincar\'e framework need not be geodesic equations.

For fluid dynamics, the Euler-Poincar\'e principle enables the derivation of equations for a fluid model from energetic interactions in a closed system. The energy is given by a Lagrangian functional on a Lie algebra. The Lagrangian is usually defined as the kinetic energy minus the potential energy. In case that the kinetic energy is positive definite and potential energy is absent, the Euler-Poincar\'e and Euler-Arnold frameworks coincide.

To state the Euler-Poincar\'e theorem, we require variational derivatives. There are different equivalent ways of defining the variational derivative. Let $J$ be a smooth functional on a Banach space $\mathcal{B}$, and let $\langle\,\cdot\,,\,\cdot\,\rangle:\mathcal{B}^*\times\mathcal{B}\to\mathbb{R}$ denote the duality pairing. The Gateaux derivative of $J$ in the direction of $\phi$ is given by
\begin{equation}
    \delta J[\psi;\phi] = \lim_{\varepsilon\to 0} \frac{J[\psi+\varepsilon\phi]-J[\psi]}{\varepsilon} =: \left\langle\frac{\delta J}{\delta\psi},\phi\right\rangle,
\end{equation}
where the right hand side defines the functional derivative $\frac{\delta J}{\delta\psi}$ as an element of $\mathcal{B}^*$. With the above definitions of the variational derivative, the statement of the Euler-Poincar\'e theorem is as follows.
\begin{theorem}[Euler-Poincar\'e]
\label{thm:EP}$\,$\\
Let $\mathfrak{X}(M)$ denote the space of vector fields on a compact finite-dimensional Riemannian manifold $M$. The following two statements are equivalent:
\begin{enumerate}
\item Hamilton's variational principle in Eulerian coordinates, with $u\in\mathfrak{X}(M)$,
\begin{equation}
\delta S := \delta\int_{t_1}^{t_2}L(u,a) \,dt= 0,
\label{eq:hvp}
\end{equation}
holds on $\mathfrak{X}(M)$, using variations of the form
\begin{equation}
\delta u = \frac{d}{dt}w - [u, w],
\end{equation}
where the arbitrary vector field $w$ vanishes at the endpoints in time.
\item The Euler-Poincar\'e equations hold. These equations are
\begin{equation}
\frac{d}{dt}\frac{\delta \ell}{\delta u} 
+ u\cdot \nabla\frac{\delta\ell}{\delta u} + \Big(\nabla\frac{\delta\ell}{\delta u}\Big)^T\cdot u + \frac{\delta\ell}{\delta u}(\nabla\cdot u)
= 0.
\label{eq:EPeq}
\end{equation}
\end{enumerate}
\end{theorem}
For the detailed statement and the proof of the Euler-Poincar\'e variational principle, see \cite{holm1998euler}. The Riemannian structure is necessary for the interpretation of the gradient $\nabla$ and the inner product $\,\cdot\,$. For the sphere, we discuss this in detail in the Appendix. An application of the Euler-Poincar\'e theorem to the Lagrangian \eqref{eq:lagrsw} yields the rotating shallow water equations on the sphere
\begin{equation}\label{eq:rsw}
    \begin{aligned}
        \frac{D}{Dt}u + \frac{1}{{\rm Ro}}fu^\perp &= -\frac{1}{{\rm Fr}^2}\nabla(h-b),\\  
        \frac{D}{Dt}h + h(\nabla\cdot u) &= 0,
    \end{aligned}
\end{equation}
where $D/Dt$ is the material derivative introduced in \eqref{eq:transport}. A detailed derivation of the rotating shallow water equations \eqref{eq:rsw} using the Euler-Poincar\'e theorem can be found in \cite{crisan2017mathematics}. It can be seen that the pressure in \eqref{eq:rsw} is hydrostatic and is generated by the dimensionless free surface elevation $\eta$, satisfying $\alpha \eta = h-b$. Let $\eta_0$ denote the typical free surface elevation, then $\alpha = \eta_0/H$ is the typical wave amplitude, which is small for geophysical flows. The material derivative does not change form when changing the dimension of the problem (going from two to three dimensions for instance), but does change form depending on coordinates. The following alternative formulation does not change when a different coordinate system is used, but changes form for different dimensions. This alternative formulation is called the vector-invariant form and is given by
\begin{equation}
\begin{aligned}
    \frac{\partial}{\partial t} u + \left(\omega+\frac{1}{{\rm Ro}}f\right)u^\perp &= -\nabla\left(\frac{1}{{\rm Fr}^2}(h-b) + \frac{1}{2}|u|^2\right),\\
    \frac{\partial}{\partial t} h + \nabla\cdot(h u) &= 0,
\end{aligned}
\label{eq:vorticity}
\end{equation}
where $\omega = \nabla^\perp\cdot u$ is the vorticity. The key identity that allows the conversion from \eqref{eq:rsw} to \eqref{eq:vorticity} is given in the Appendix by \eqref{eq:fvifd}.  Note that the vorticity is scalar valued, this is one of the key differences between two-dimensional and three-dimensional fluid dynamics. In what follows, we consider only the vector-invariant form of the rotating shallow water equations. By a careful analysis of the scales involved in typical geophysical problems, the dominant contributions to the dynamics can be identified and the rotating shallow water equations can be simplified. This is considered next.

\subsection{Conventional geostrophic balance}
For flows at planetary scales, the Rossby number and the Froude number are both small and usually of similar magnitude. The similarity between the Rossby number and the Froude number is measured by the Burger number $\mathrm{Bu} = (\mathrm{Ro}/\mathrm{Fr})^2$. A small value for the Burger number indicates that rotation dominates, while a large value indicates that the flow is dominated by stratification. In Table \ref{tab:scales} below, typical parameter values of mid-latitude planetary flows are given for three examples: atmospheric flows on Earth, oceanic flows on Earth and atmospheric flows on Jupiter. 
\begin{table}[H]
    \centering
    \begin{tabular}{r|r|r|r}
    Flow & Atmospheric (Earth) & Oceanic (Earth) & Atmospheric (Jupiter) \\[5pt]
    \hline & & \\[-5pt]
    Length $L$ & $10^7$ m & $3\cdot 10^6$ m  & $10^8$ m \\[5pt]
    Height $H$ & $2.5\cdot 10^3$ m & $10^2$ m & $1.5\cdot 10^5$ m \\[5pt]
    Velocity $U$ & $10^1$ m/s & $ 10^{-1}$ m/s & $10^2$ m/s \\[5pt]
    Wave height $\eta_0$  & $6\cdot 10^2$ m & $3\cdot 10^0$ m & $5\cdot 10^3$ m\\[5pt]
    Rotation frequency $\Omega$ & $1.2 \cdot 10^{-5}$ Hz & $1.2 \cdot 10^{-5}$ Hz & $2.8 \cdot 10^{-5}$ Hz \\[5pt]
    Gravitational acceleration $g$ & 9.8 m/s$^2$ & 9.8 m/s$^2$ & 24.8 m/s$^2$ \\[5pt]
    \hline & & & \\[-5pt]
    Rossby number $\mathrm{Ro}$ & $6.1\cdot 10^{-2}$ & $4.0 \cdot 10^{-3}$ & $5.0\cdot 10^{-2}$\\[5pt]
    Froude number $\mathrm{Fr}$ & $6.4\cdot 10^{-2}$ & $3.2 \cdot 10^{-3}$ & $5.2\cdot 10^{-2}$\\[5pt]
    Wave amplitude $\alpha$ & $6\cdot 10^{-2}$ & $3\cdot 10^{-3}$ & $5\cdot 10^{-2}$\\[5pt]
    \hline & & & \\[-5pt]
    Burger number $\mathrm{Bu}$ & 0.91 & 1.56 & 0.92
    \end{tabular}
    \caption{Typical length scales, velocity scale, planetary rotation frequency and planetary gravitational acceleration for flows on Earth and Jupiter, and the values for the Rossby number, Froude number and wave amplitude that these scales imply. The final row contains the Burger number.}
    \label{tab:scales}
\end{table}

With the scales as above, it can be noted that the Rossby number, the Froude number and the wave amplitude are all small and of similar order of magnitude. This is a characteristic feature of geophysical flows at large scales. When the three dimensionless numbers $\mathrm{Ro}, \mathrm{Fr}, \alpha$ are of similar order of magnitude, i.e., $\mathcal{O}(\mathrm{Ro}) = \mathcal{O}(\mathrm{Fr}) = \mathcal{O}(\alpha)$, the gradient of the Coriolis parameter satisfies $\nabla f = \mathcal{O}(\mathrm{Ro})$ and the bottom topography satisfies $\nabla h = \mathcal{O}(\mathrm{Ro})$, then the conditions for geostrophic balance are satisfied. 

Since all small parameters are of similar order of magnitude, we can introduce a single small parameter $\varepsilon$ that represents all of these small physical parameters. This is convenient for the computations since it becomes simpler to group terms are of similar size in the equations. However, to establish the parameter regimes of the physical parameters, it remains convenient to keep track of the parameters $\alpha,{\rm Ro}, {\rm Fr}$ in the equations. We can then introduce regular perturbation series for the variables and parameters in the shallow water equations as follows
\begin{equation}
    \begin{aligned}
        u &= u_0 + \varepsilon u_1 + \varepsilon^2 u_2 + \mathcal{O}(\varepsilon^3),\\
        \eta &= 1 + \varepsilon \eta_1 + \varepsilon^2 \eta_2 + \mathcal{O}(\varepsilon^3),\\
        f &= f_0 + \varepsilon f_1 + \varepsilon^2 f_2 +\mathcal{O}(\varepsilon^3),\\
        \zeta &= \varepsilon\zeta_1 + \varepsilon^2\zeta_2 + \mathcal{O}(\varepsilon^3),\\
        h &= \varepsilon h_1 + \varepsilon^2 h_2 + \mathcal{O}(\varepsilon^3).
    \end{aligned}
    \label{eq:perturbationseries}
\end{equation}
We have used the fact that in flows of geophysical scale the perturbation series expansions of $\zeta$ and $h$ start at order $\mathcal{O}(\epsilon)$. In \eqref{eq:perturbationseries}, we have an expansion of the Coriolis parameter. The most important component of this expansion is $f_0$, which in $f$-plane and $\beta$-plane formulations relates to the constant latitude at which one considers the tangent plane. In the $\beta$-plane case, one further has the explicit expression $f_1=y$ in which $y$ denotes the vertical coordinate in a tangent plane. Upon substituting these perturbation series expansions into \eqref{eq:vorticity}, one finds at leading order $\mathcal{O}(\varepsilon^{-1})$
\begin{equation}\label{eq:geobal}
    f_0 u_0^\perp + \nabla\zeta_1 = 0.
\end{equation}
This is the equation for geostrophic balance. So we identify the leading order component of the velocity $u_0$ with the geostrophic velocity field that satisfies the above equation. One can proceed by deriving a closed system of equations at the next order, which describes a perturbation around geostrophic balance. Establishing a closed model at this stage leads to the quasi-geostrophic equations, but only away from the equator, since on the equator $f_0=0$ and \eqref{eq:geobal} is not defined. Since $\nabla f_0=\mathcal{O}(\mathrm{Ro})=\mathcal{O}(\varepsilon)$, the leading order velocity field $u_0$ is divergence-free by the vector calculus identities that we introduced in Section \ref{sec:spherical}. 

To derive a global quasi-geostrophic model on the sphere, planar approximations cannot be used since $f_0=0$ at the equator. We therefore present an alternative route to the quasi-geostrophic equations on the sphere in the next subsection, avoiding the problems at the equator.

\subsection{Geostrophic balance using vorticity and divergence variables}
An alternative approach to the conventional derivation of the geostrophic balance is to first express the rotating shallow water equations in terms of vorticity and divergence variables. We obtain the vorticity equation by applying the perpendicular divergence to the velocity equation in \eqref{eq:rsw} and we obtain the equation governing the evolution of the divergence $D=\nabla\cdot u$ by applying the divergence operator to the velocity equation \eqref{eq:rsw}. Together with the continuity equation, this yields the closed system of equations given by
\begin{equation}
\begin{aligned}
    \frac{\partial}{\partial t} \omega + \nabla\cdot\left(\Big(\omega + \frac{1}{{\rm Ro}}f\Big)u\right) &= 0,\\
    \frac{\partial}{\partial t}D + \nabla\cdot\left(\Big(\omega + \frac{1}{{\rm Ro}}f\Big)u^\perp\right) &= - \Delta\left(\frac{1}{{\rm Fr}^2}(\eta-h) + \frac{1}{2}|u|^2\right)\\
    \frac{\partial}{\partial t} \eta + \nabla\cdot(\eta u) &= 0.
\end{aligned}
\label{eq:rswvordiv}
\end{equation}
The system of equations \eqref{eq:rswvordiv} is completely equivalent to the rotating shallow water equations in \eqref{eq:rsw}. Note that all equations in \eqref{eq:rswvordiv} take the form of hyperbolic conservation laws. The right-hand side of the divergence equation shows that divergence is not just transported, but also created or annihilated. Indeed, the right-hand side of the divergence equation contains the source terms for inertia-gravity waves. It can now be shown that the potential vorticity, which is defined by 
\begin{equation}\label{eq:pv}
    q_{rsw} = \frac{1}{\eta}\left(\omega + \frac{1}{{\rm Ro}}f \right)
\end{equation}
is a Lagrangian invariant for the rotating shallow water equations, meaning that it satisfies the equation
\begin{equation}\label{eq:pvevolution}
    \frac{D}{Dt}q_{rsw} = 0.
\end{equation}
The potential vorticity gives rise to an infinite family of Casimirs, which are integral invariants of the form
\begin{equation}
    C_\Phi = \int_{S^2} \Phi(q_{rsw})\,\eta\,dA,
\end{equation}
where $\Phi$ is any analytic function. Taking $\Phi(q_{rsw})=q_{rsw}^2$, one obtains the potential enstrophy, which, since it is a Casimir, is a conserved quantity of the rotating shallow water equations. It is different from the usual enstrophy variable that one encounters in incompressible fluids since the mass density is not constant. The presence of infinitely many Casimirs motivates approximating the potential vorticity rather than the velocity directly. Let us recall the Hodge decomposition \eqref{eq:hodge} of the velocity field, i.e., we introduce the stream function $\psi$ and velocity potential $\chi$ such that $u = \nabla^\perp\psi + \nabla\chi$. 

The fact that the leading order term $u_0$ of the velocity is divergence free at leading order in $\epsilon$ motivates an adaptation of the Hodge decomposition. In fact, we explicitly impose that the leading order velocity field in the regime of geostrophic balance is nearly divergence free,
\begin{equation}\label{eq:hodgedecomp}
u_0 = \nabla^\perp \psi_0 + \varepsilon\nabla\chi_1.
\end{equation}
This weighted decomposition shows that the divergence variable obeys $D=\epsilon\Delta\chi$. Inserting the decomposition into the divergence equation in \eqref{eq:rswvordiv}, using the fact that all small parameters are of similar size yields after some rewriting
\begin{equation}
    \epsilon\frac{\partial}{\partial t}\Delta\chi_1 + \nabla\cdot\left(-\Big(\omega+\frac{1}{\varepsilon}f_0\Big)\nabla\psi_0 + \frac{1}{\varepsilon}\nabla\zeta_1 + \frac{1}{2}|\nabla\psi_0|^2\right) + \mathcal{O}(\epsilon)= 0.
\end{equation}
The leading order terms in this equation yield the divergence of the geostrophic balance condition that we derived in the previous section. Namely, at order $\mathcal{O}(\epsilon^{-1})$, we have 
\begin{equation}\label{eq:bal}
    \nabla\cdot\left(-f_0\nabla\psi_0 + \nabla\zeta_1\right) = 0.
\end{equation}
Equation \eqref{eq:bal} is known as the linear balance equation and goes back to the work of \cite{lorenz1960energy}. Since by assumption $\nabla f_0 = \mathcal{O}(\varepsilon)$, we can further simplify the above equation to obtain
\begin{equation}
    \Delta(f_0\psi_0 + \zeta_1) = 0.
\end{equation}
Upon retracing the steps of the computations up to now with the physical parameters in place instead of $\varepsilon$, one obtains the following expression
\begin{equation}
    \Delta\left(-f_0\psi_0 + \frac{\alpha\,{\rm Bu}}{{\rm Ro}}\zeta_1\right) = 0,
\end{equation}
where the ratio $\alpha\,{\rm Bu}\,{\rm Ro}^{-1}$ in front of the free surface elevation is of order $\mathcal{O}(1)$, since ${\rm Bu} = \mathcal{O}(1)$ and $\mathcal{O}(\alpha)=\mathcal{O}({\rm Ro})$. The trivial solution to this equation is the simplest form of a geostrophic relation, \cite{daley1983linear}, and is given by
\begin{equation}\label{eq:spheregeobal}
f_0\psi_0 = \frac{\alpha\,{\rm Bu}}{{\rm Ro}}\zeta_1.
\end{equation}
Using the relation \eqref{eq:spheregeobal}, the potential vorticity \eqref{eq:pv} can now be approximated. The layer thickness $\eta$ in nondimensional form is written as
\begin{equation}\label{eq:depth}
    \eta = 1 + \alpha\zeta_1 - \alpha h_1 + \mathcal{O}(\alpha^2),
\end{equation}
which also dictated the perturbation series expansion that was introduced in \eqref{eq:perturbationseries}. This shows that $\eta = 1+\mathcal{O}(\alpha)$ which allows the approximation of $1/\eta$ in the following way
\begin{equation}
    \frac{1}{\eta} = 1-\alpha\zeta_1+\alpha h_1 + \mathcal{O}(\alpha^2).
\end{equation}
It can be checked that with this approximation the identity $1 = \eta (1/\eta)$ holds up to $\mathcal{O}(\alpha)$. We can now expand the potential vorticity \eqref{eq:pv} as
\begin{equation}
    q_{rsw} = \left(\omega_0 + \frac{1}{{\rm Ro}}f_0 - \alpha\zeta_1\omega_0 - \frac{\alpha}{{\rm Ro}} f_0\zeta_1 + \alpha h_1\omega_0 + \frac{\alpha}{{\rm Ro}}f_0 h_1\right) + \mathcal{O}(\alpha^2).
\end{equation}
We define the quasi-geostrophic potential vorticity $q$ to be 
\begin{equation}
    q_{rsw} = q + \mathcal{O}(\alpha) = \frac{1}{{\rm Ro}}f_0 + \omega_0 - \frac{\alpha}{{\rm Ro}}f_0\zeta_1 + \frac{\alpha}{{\rm Ro}}f_0 h_1 + \mathcal{O}(\alpha).
\end{equation} 
The terms $\alpha\zeta_1\omega_0$ and $\alpha h_1\omega_0$ are both of order $\mathcal{O}(\alpha)$ and are therefore absorbed in the symbol $\mathcal{O}(\alpha)$. By means of \eqref{eq:spheregeobal} and the relation $\Delta\psi_0 = \omega_0$ we obtain the relation between the potential vorticity and the stream function. We now have obtained a closed model. At this moment we discard terms of higher order and no longer track the subscripts that come from the asymptotic analysis. This means that from here onward, the model stands alone as
\begin{equation}\label{eq:qgpv}
\begin{aligned}
    q &= \frac{1}{{\rm Ro}}f + \Delta\psi - \frac{1}{{\rm Bu}}f^2\psi + \frac{\alpha}{{\rm Ro}}f h.
\end{aligned}
\end{equation}
An alternative form is given in terms of Lamb's parameter $\gamma = 4\Omega^2 a^2/(gH) = a^2/(L^2\,{\rm Bu})$, which can be viewed as the square of the ratio of the earth's radius $a$ and the Rossby deformation radius $gH/f^2$. In the setting with the Lamb parameter, vorticity and potential vorticity are measured in units of $\Omega^{-1}$, the velocity in units $a\Omega$ and the height variations $\zeta$, $h$ in terms of the average layer thickness $H$. Letting $\mu = \cos \theta$, the potential vorticity is expressed as
\begin{equation}\label{eq:qgalternative}
    q = \frac{2\mu}{{\rm Ro}} + \Delta\psi - \gamma\mu^2\psi + 2\mu h,
\end{equation}
which is the dimensionless version of the expressions found in \cite{verkley2009balanced,schubert2009shallow}. The quasi-geostrophic potential vorticity on the sphere features a factor $f^2$ (or $\mu^2$, depending on the convention). This means that the elliptic relation between the potential vorticity and the stream function for the quasi-geostrophic model on the sphere is an inhomogeneous Helmholtz operator
\begin{equation}
    q = (\Delta - \gamma\mu^2)\psi + \frac{2\mu}{{\rm Ro}} + 2\mu h.
    \label{eq:pvinversion}
\end{equation}
In the planar case, the relation between the potential vorticity and the stream function is given by a homogeneous Helmholtz operator. The equation that governs the evolution of potential vorticity takes the same form as \eqref{eq:pvevolution}, though it is commonly expressed by means of the stream function as
\begin{equation}\label{eq:qgsphere}
    \frac{\partial}{\partial t}q + \{\psi,q\} = 0,
\end{equation}
using the relation between transport and the Poisson bracket elucidated in \eqref{eq:transportandpb}. The vorticity-stream function formulation is particularly convenient because one does not need to compute the pressure. Instead, one solves the elliptic problem \eqref{eq:pvinversion}. The bracket in \eqref{eq:qgsphere} is the canonical Poisson bracket of $C^\infty(S^2)$ functions \eqref{eq:transportandpb}, which appears naturally for two-dimensional incompressible flows, as shown by \cite{marsden1983coadjoint}. It serves as an indicator of the rich geometric structure that one finds in two-dimensional incompressible ideal fluids and motivates suitable structure-preserving numerical methods. We will now show that the quasi-geostrophic equations are a Hamiltonian system with respect to the above Poisson bracket and have infinitely many conserved integral quantities.

\section{Lagrangian and Hamiltonian formulation}\label{sec:lagrangian}
The key to the derivation of quasi-geostrophy on the sphere is the small divergence of the velocity field. This implies that the fluid dynamics at planetary scales are completely determined by the potential vorticity and the stream function, which implies that the flow determined by the equations is incompressible. For incompressible fluids in two dimensions, the kinetic energy $K$ is given by the integral over the domain of $-\frac{1}{2}q\psi$, see \cite{marsden1983coadjoint}. Together with the relations $u=\nabla^\perp\psi$ and \eqref{eq:pvinversion}, we can formulate the kinetic energy in terms of the velocity by integration by parts. Here we use the relations provided by the De Rham complex and find
\begin{equation}
    \begin{aligned}
        K &= \int_{S^2} -\frac{1}{2}q\psi dA \\
        &= \int_{S^2} -\frac{1}{2}\psi(\Delta-\gamma\mu^2)\psi - \frac{\mu}{{\rm Ro}}\psi - \mu h\psi\, dA\\
        &= \int_{S^2} \frac{1}{2}\nabla^\perp \psi \cdot\nabla^\perp\psi + \frac{1}{2}\nabla^\perp\psi (\gamma\mu^2\Delta^{-1})\nabla^\perp\psi - \left(\frac{\mu}{{\rm Ro}} + \mu h\right)\psi\, dA\\
        &= \int_{S^2} \frac{1}{2}u\cdot(1-\gamma\mu^2\Delta^{-1})u + u\cdot V\,dA
    \end{aligned}
\end{equation}
where $\nabla^\perp\cdot V = -({\rm Ro}^{-1}\mu +\mu h)$. Note that $\nabla^\perp\psi\cdot\nabla^\perp\psi$ is the length of the rotated gradient of $\psi$. Since rotation preserves lengths, this is the same as the length of the gradient of $\psi$. The interpretation of the kinetic energy in terms of the velocity indicates that the quasi-geostrophic model has a different metric than the 2D Euler equations. Indeed, the 2D Euler equations are obtained by minimising the $L^2$ kinetic energy, see \cite{arnold1966principe}, whereas the kinetic energy corresponding to the quasi-geostrophic equations follow from an inhomogeneous $H^{-1}$-metric. The term in addition to the $L^2$ kinetic energy in the quasi-geostrophic equations is multiplied by the Lamb parameter $\gamma$ and is associated with Cressman stretching (see \cite{verkley2009balanced} and \cite{cressman1958barotropic}). Geometrically, this term changes the sphere to an oblate spheroid. The oblateness of the spheroid depends on how fast the sphere rotates, which is reflected by the factor $\mu^2$. The vector field $V$ is the vector potential that produces the effect of the rotating frame and the nontrivial bottom topography. The kinetic energy is used to define the Lagrangian and the Hamiltonian.

\subsection{Energy functionals}
The Lagrangian should be formulated such that the potential vorticity and the stream function can determine the dynamics completely, i.e., there is no velocity potential. This is equivalent to an incompressibility constraint \cite{holm1998euler}, which we introduce in the Lagrangian with a Lagrange multiplier $p$ that plays the role of a pressure. This is necessary because the variational point of view describes the relation between velocity and momentum, and at the moment we have not introduced a momentum variable yet. The resulting Lagrangian consists of the kinetic energy with respect to a metric that encodes the relation \eqref{eq:pvinversion} and an incompressibility constraint
\begin{equation}\label{eq:lagqg}
    L(u,\eta) = \int_{S^2} \frac{1}{2}\eta u\cdot (1-\gamma \mu^2\Delta^{-1}) u + \eta u\cdot V - p(\eta-1)\,dA,
\end{equation}
where $\eta u$ is the momentum variable. The Lagrangian \eqref{eq:lagqg} is closely related to the Lagrangian for planar quasi-geostrophy proposed in \cite{holm1998hamilton}, which was derived using the Lagrangian perspective. The key difference is that \eqref{eq:lagqg} includes an elliptic operator that depends on the full Coriolis parameter, extending the previous work in~\cite{holm1998hamilton} and that the present work uses the Eulerian interpretation. At this stage, applying the Euler-Poincar\'e theorem to the above Lagrangian yields the quasi-geostrophic equations on the sphere in velocity form. However, we prefer the equations in terms of potential vorticity and stream function, since this expresses the model completely in terms of scalar functions. By representing the velocity field as $u=\nabla^\perp \psi$, which makes the incompressibility constraint redundant, we can formulate the unconstrained Lagrangian $L(\psi)$ as
\begin{equation}
\begin{aligned}
    L(\psi) &= \int_{S^2} \frac{1}{2}\nabla^\perp \psi \cdot (1 - \gamma\mu^2\Delta^{-1})\nabla^\perp \psi + \nabla^\perp\psi\cdot V \,dA \\
    &= \int_{S^2} \frac{1}{2}|\nabla \psi|^2 + \frac{1}{2}\gamma\mu^2 \psi^2 + \left(\frac{1}{\epsilon}\mu + \mu h\right)\psi \,dA.
\end{aligned}
\end{equation}
Since we now have an unconstrained Lagrangian, we can apply the Legendre transformation to obtain the Hamiltonian. We define the dual variable to the stream function equal to the variational derivative of the Lagrangian with respect to the stream function and compute the variational derivative of the Lagrangian with respect to the stream function
\begin{equation}
    \frac{\delta L}{\delta \psi} = (\gamma\mu^2-\Delta)\psi + \frac{1}{\epsilon}\mu + \mu h = q,
\end{equation}
which is precisely the potential vorticity \eqref{eq:qgalternative}. The Legendre transform then defines the Hamiltonian $H(q)$ as
\begin{equation}\label{eq:hamiltonian}
\begin{aligned}
    H(q) &= \int_{S^2} q\psi\, d\mu - L(\psi) = \int_{S^2} \frac{1}{2}|\nabla\psi|^2 + \frac{1}{2}\gamma\mu^2\psi^2\, d\mu\\
    &= \int_{S^2} \left(q-\frac{1}{\epsilon}\mu - \mu h\right)(\Delta-\gamma\mu^2)^{-1}\left(q-\frac{1}{\epsilon}\mu - \mu h \right)\, dA
\end{aligned}
\end{equation}
Equality of the Lagrangian and Hamiltonian occurs for pure geodesic problems (see \cite{arnold1966geometrie}), which is the category that flows of incompressible ideal fluids belong to. The linear terms that arise due to rotation and bottom topography constitute the difference between the Lagrangian and Hamiltonian for quasi-geostrophic flow on the sphere. These terms are obtained from the relation between the stream function and the potential vorticity. From a mathematical point of view, these linear terms are in the center of the Poisson algebra $C^\infty(S^2)$, since they do not depend on time, but do influence the dynamics. This geometric insight is valuable and \cite{modin2025geodesic} derive a stability criterion based the Lamb parameter using techniques of global analysis.

The Hamiltonian \eqref{eq:hamiltonian} together with a Lie-Poisson bracket produces the dynamics. In the next section we introduce the Lie-Poisson bracket and show how one can identify the Jacobi-Lie bracket of incompressible vector fields on $S^2$ with the Poisson bracket on $C^\infty(S^2)$. One of the key implications of such a formulation for the spherical quasi-geostrophic equations is that they admit point vortex solutions.

\subsection{Geometry of two-dimensional fluids}
The geometry of two-dimensional incompressible ideal fluids was described in detail in \cite{marsden1983coadjoint} for any smooth two-dimensional manifold, including the sphere. We repeat some of the arguments here to provide a link with the numerical discretisation we use to simulate incompressible fluid dynamics on the sphere. 

The central fact that connects incompressible fluids with symplectic geometry is the following. In two dimensions, a volume element is also a symplectic structure, so that each divergence-free vector field $u$ can be thought of as a Hamiltonian vector field $X_\psi$. The stream function $\psi$ serves as a Hamiltonian and because the sphere is connected, $\psi$ is determined up to a constant by $u$. It follows that it is possible to identify the Lie algebra $\mathfrak{X}_{vol}$ of divergence-free vector fields with the smooth functions on the sphere modulo the constants $C^\infty(S^2)/$const. The dual space $\mathfrak{X}_{vol}^*$ is then identified with generalised functions $q$ on $S^2$ satisfying $\int_{S^2} q dA=0$. 

The Lie algebra bracket $[\,\cdot\,,\,\cdot\,]_{\mathfrak{X}_{vol}}$ on $\mathfrak{X}_{vol}$ is minus the Jacobi-Lie bracket, i.e., the usual commutator of vector fields:
\begin{equation}
    [u_1,u_2]_{\mathfrak{X}_{vol}} = -\mathcal{L}_{u_1} u_2 = -\mathcal{L}_{X_{\psi_1}} X_{\psi_2} = X_{\{\psi_1,\psi_2\}},
\end{equation}
which means that the Lie algebra bracket corresponds to the Poisson bracket $\{\psi_1,\psi_2\}$ of stream functions. This identification is valid for any two-dimensional connected smooth manifold, hence also applicable to the sphere. The difference in sign between the Lie bracket and Jacobi-Lie bracket on the Lie algebra of divergence-free vector fields is caused by the fact that in fluid dynamics the correspondence between the Lagrangian particle formulation and the Eulerian field formulation relies on right-invariance, whereas the Jacobi-Lie bracket is for left-invariant vector fields. The Lie-Poisson bracket $\{\,\cdot\,,\,\cdot\,\}_{LP}$ on the sphere is given by
\begin{equation}\label{eq:pbq}
    \{F,G\}_{LP}(q) = \int_{S^2} q\left\{\frac{\delta F}{\delta q},\frac{\delta G}{\delta q}\right\} dA, \qquad \forall F,G:C^\infty(S^2)\to\mathbb{R},
\end{equation}
where the Poisson bracket in coordinates is given in \eqref{eq:transportandpb}. The variational derivatives $\delta F/\delta q$ and $\delta G/\delta q$ in \eqref{eq:pbq} are interpreted as functions on $S^2$, whereas $F, G$ are functionals $C^\infty(S^2)\mapsto\mathbb{R}$. Note the cancellation of the metric factor in the measure $d\mu$ with the factor in front of the Poisson bracket \eqref{eq:transportandpb}, which shows that the Lie-Poisson bracket is coordinate invariant. The Lie-Poisson bracket \eqref{eq:pbq} has a kernel consisting of functionals $C_\Phi:C^\infty(S^2)\to\mathbb{R}$ with the expression
\begin{equation}
    C_\Phi(q) = \int_{S^2} \Phi(q) dA,
\end{equation}
where $\Phi$ is any analytic function. This means that the kernel consists of uncountably infinite many functionals. In particular, this means that any integrated power of the (potential) vorticity is conserved by the flow, i.e., for the monomials $C_n = \int_{S^2} q^n dA$ we have
\begin{equation}
    \{C_n, G\}_{LP} = 0, \qquad \forall G:C^\infty(S^2)\to\mathbb{R}.
\end{equation}
The functionals $C_\Phi$ are called Casimirs and they are ubiquitous in physics. A particularly important member of the monomial Casimir family is the enstrophy $C_2$, which plays a fundamental role in the double cascade predicted by \cite{kraichnan1967inertial} in two dimensional fluid dynamics. The quasi-geostrophic equations on the sphere with forcing and dissipation also have a double cascade, as shown by \cite{franken2023zeitlin}. The reason that there is no profound difference between Casimirs that are functionals of potential vorticity and Casimirs that are functionals of the ordinary vorticity is the following. The relation between the stream function $\psi$ and the vorticity $\omega$ is given by a Poisson equation. Changing the Poisson equation to the inhomogoneous Helmholtz equation corresponds to a change of metric, which does not affect the topology of the system. Secondly, the inclusion of rotation corresponds to a central extension of the Lie algebra, see for instance \cite{zeitlin1994differential}. A central extension corresponds to adding an extra vector to the basis of the Lie algebra such that this new vector commutes with all other basis elements (making the extension central). At the same time, the extension appears nontrivially in the commutators of the other basis elements. This means that the new vector does not increase the dimension defined by the set of dynamical variables, but does influence the dynamics. This is how the Coriolis parameter affects rotating incompressible ideal fluids in two dimensions. For the sphere it can be explicitly shown that the Coriolis parameter corresponds to a central extension by using the basis of spherical harmonics as in \cite{zeitlin2004self}. 

With $H$ the Hamiltonian given by \eqref{eq:hamiltonian} and the Lie-Poisson bracket \eqref{eq:pbq}, the quasi-geostrophic equations on the sphere are a Lie-Poisson Hamiltonian system. For any observable $F:C^\infty(S^2)\to\mathbb{R}$ of potential vorticity, we have
\begin{equation}
    \frac{d}{dt}F(q) = \{F(q),H(q)\}_{LP}.
\end{equation}
The quasi-geostrophic equations on the sphere are given by
\begin{equation}\label{eq:qg}
    \frac{\partial}{\partial t}q + \left\{\Big(\Delta - \gamma\mu^2\Big)^{-1}\Big(q-\frac{1}{\epsilon}\mu-\mu h\Big) , q\right\} = 0,
\end{equation}
where the first term in the Poisson bracket is the stream function. Thus we have derived \eqref{eq:qgsphere} from a geometric point of view. The benefit of the geometric derivation is that it establishes uncountably many invariants. While this infinite family of Casimirs is present for any simply connected two-dimensional smooth manifold, in the case of the sphere powerful numerical methods are have been recently developed, making large-scale simulation of the quasi-geostrophic equations on the sphere possible. This is what we turn to next. These numerical methods use the Lie-Poisson formulation in combination with the isospectrality property that is associated with the domain being the sphere. These isospectral Lie-Poisson solvers were introduced in \cite{modin2020casimir} and are able to preserve $N-1$ monomial Casimirs to machine precision when the number of basis functions is $N^2$. Having shown that the quasi-geostrophic equations on the sphere are a Lie-Poisson system, we can use isospectral Lie-Poisson solvers to integrate \eqref{eq:qg} guaranteeing the preservation of Casimirs.

\section{Structure-preserving numerical simulation of quasi-geostrophic dynamics on a rotating sphere}\label{sec:numerics}
In this section, we elaborate on numerical simulations of quasi-geostrophy on the sphere without viscous dissipation. For the hyperbolic version of the quasi-geostrophic equations, numerical simulation is particularly challenging due to potential numerical errors that break the geometric structure that the equations possess. As far as we know at present, there is a single numerical method that is able to preserve energy as well as Casimir invariants, which is the aforementioned isospectral Lie-Poisson solver. This method works as follows.

Spherical harmonics are the eigenfunctions of the spherical Laplacian and provide an orthonormal basis on the sphere. Upon replacing the spherical Laplacian by a discrete $N$-Laplacian that share eigenvalues for the first $N$-monomial functions, the Poisson bracket can be truncated in a self-consistent way, see \cite{zeitlin2004self}. This truncation yields a finite dimensional system that retains the Lie-Poisson structure and converges to \eqref{eq:qg} as the number of modes is increased to infinity, see \cite{hoppe1989diffeomorphism}. For the details of the numerical method we refer to \cite{modin2020casimir, cifani2023efficient}. In \cite{cifani2022casimir}, the isospectral Lie-Poisson integrator was used to simulate two-dimensional turbulence and strong numerical evidence for the double cascade was provided, indicating an accurate numerical representation of turbulent interactions in the flow. In \cite{franken2023zeitlin} an in-depth analysis of the implication of this method for the double cascade mechanism and the Rhines barrier in the quasi-geostrophic equations with forcing and dissipation can be found. Here we focus on the hyperbolic case of the quasi-geostrophic equations given in \eqref{eq:qg} without external forcing or dissipation. 

Since the numerical method preserves the Lie-Poisson structure, quasi-geostrophic flows can be simulated in the absence of numerical dissipation, which allows us to study freely developing solutions on the sphere. Here, we demonstrate this property by simulating the emergence of coherent large-scale flow structures from initial small-scale vortices. We simulate the quasi-geostrophic equations on a unit sphere with trivial bottom topography, rotating at 20 revolutions per second. The Lamb parameter is set to $\gamma=10^3$. The total energy in the system is determined by the initial condition, and is tuned such that the horizontal velocity scale is $\mathcal{O}(1)$. This leads to a Rossby number of $\mathcal{O}(10^{-2})$, thus ensuring that the flow is in the geostrophic regime. 

The numerical method is based on a truncation of spherical harmonics, which is set at $N=512$ for this experiment, meaning that modes up to degree $l=511$ are resolved. The initial condition is completely determined by the potential vorticity $q$, and is chosen such that energy is only contained in small length scales. Specifically, we select only modes with degree $40<l<60$, which are given a fixed amplitude $A_{0}$ and a random phase. Using some preliminary simulations, the amplitudes are set to $A_{0}=1/50$ to ensure horizontal velocity scales of order unity.

\begin{figure}[H]
    \centering
    \begin{subfigure}{.47\textwidth}\centering
        \includegraphics[width=\columnwidth]{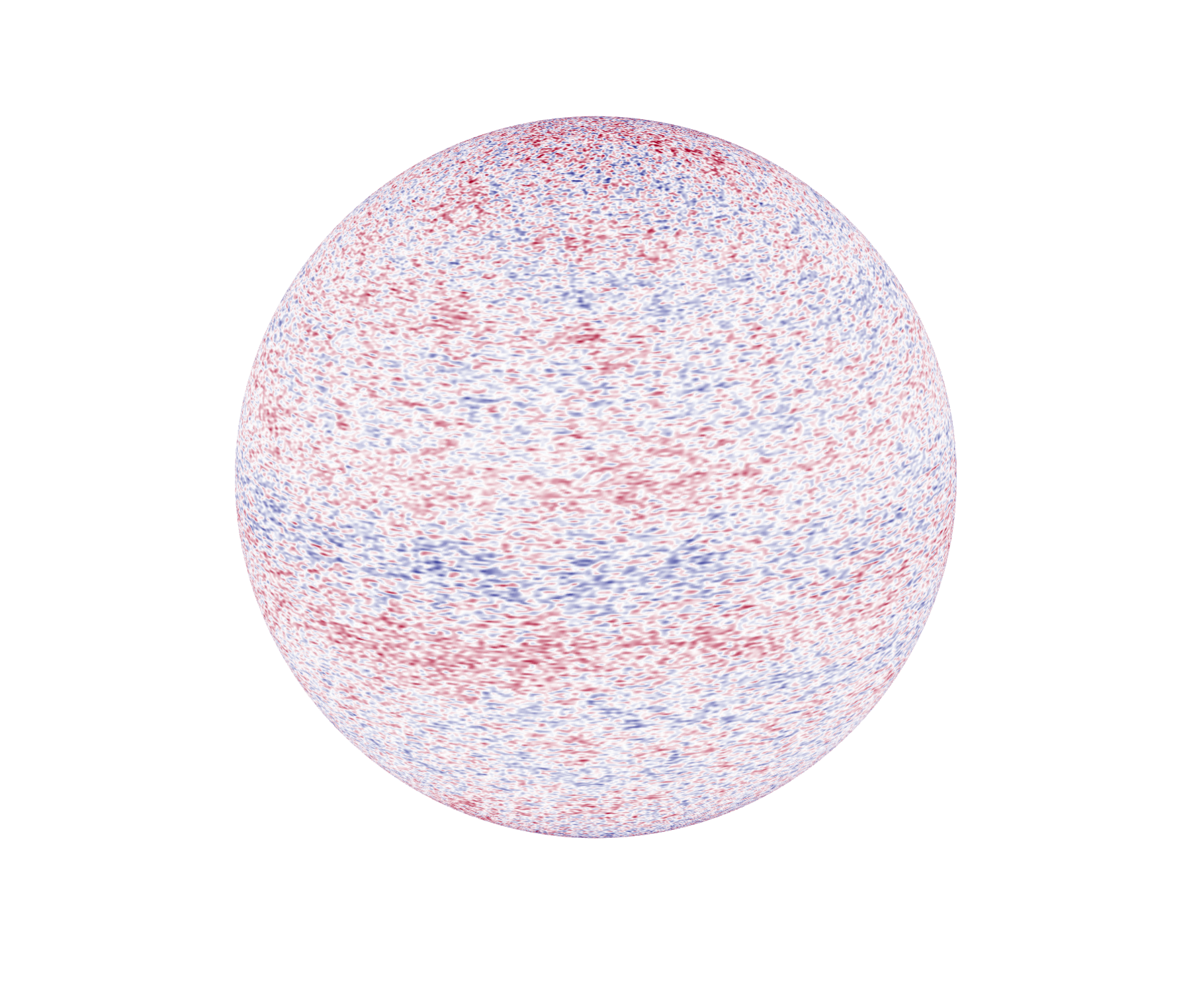}
    \end{subfigure}
    \begin{subfigure}{.47\textwidth}\centering
        \includegraphics[width=\columnwidth]{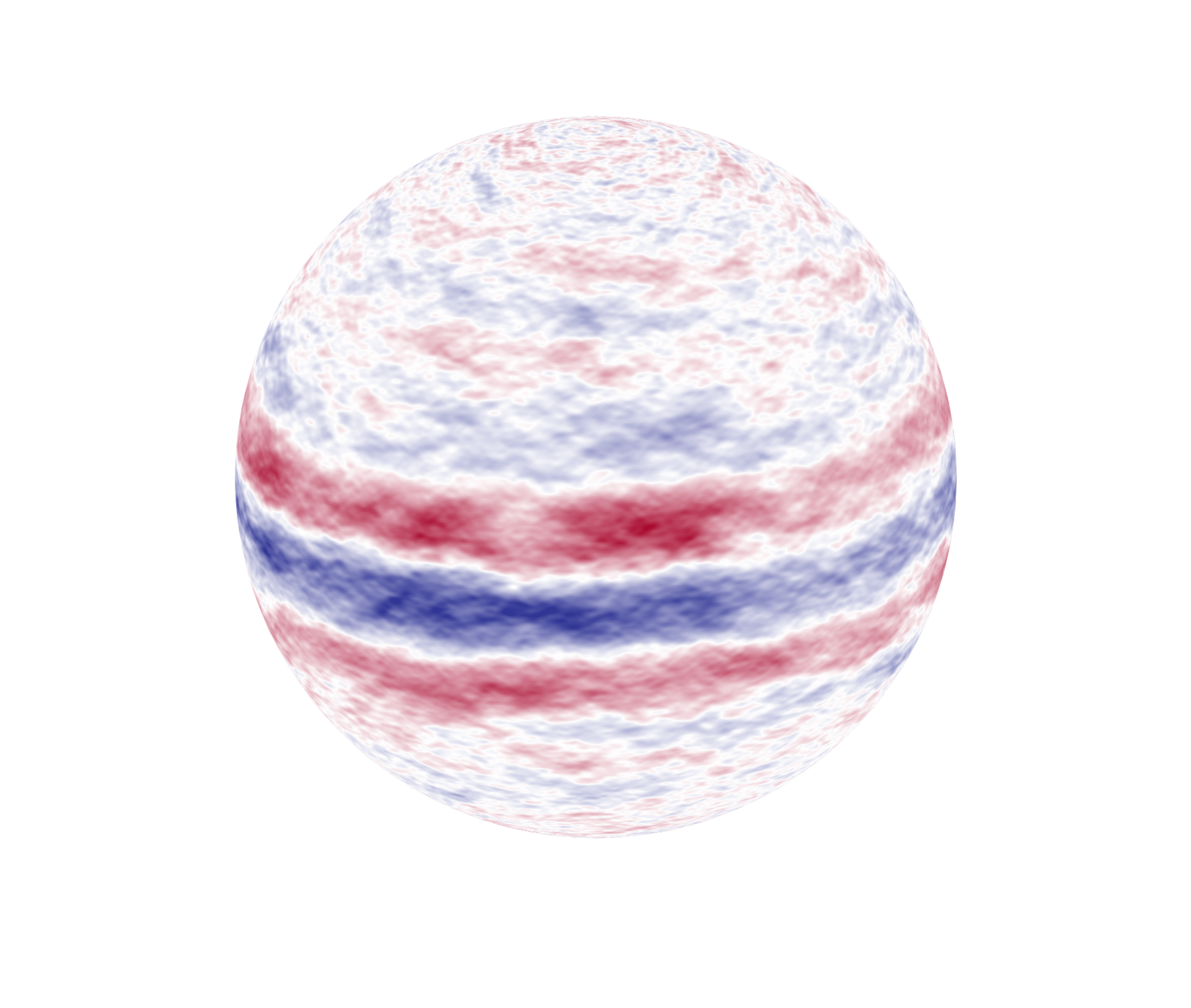}
    \end{subfigure}
    \caption{Potential vorticity anomaly (left) and zonal component of the velocity field (right) after 2000 days.}
    \label{fig:spheres}
\end{figure}

We simulate the flow of potential vorticity for 2000 days at a time step of 1/125 days. Figure~\ref{fig:spheres} shows the potential vorticity anomaly $q-\frac{1}{\epsilon}\mu$, i.e., the potential vorticity in a co-rotating reference frame of the sphere, and the zonal velocity after 2000 days. The zonal velocity is calculated from the gradient of the streamfunction, which in turn can be calculated from the potential vorticity at any time using equation \ref{eq:pvinversion}. From the zonal velocity, we clearly see the emergence of zonal jets in the equatorial region as alternating coherent flow structures in the longitudinal direction. 

\begin{figure}
    \centering
    \begin{subfigure}{.47\textwidth}\centering
        \includegraphics[width=\columnwidth]{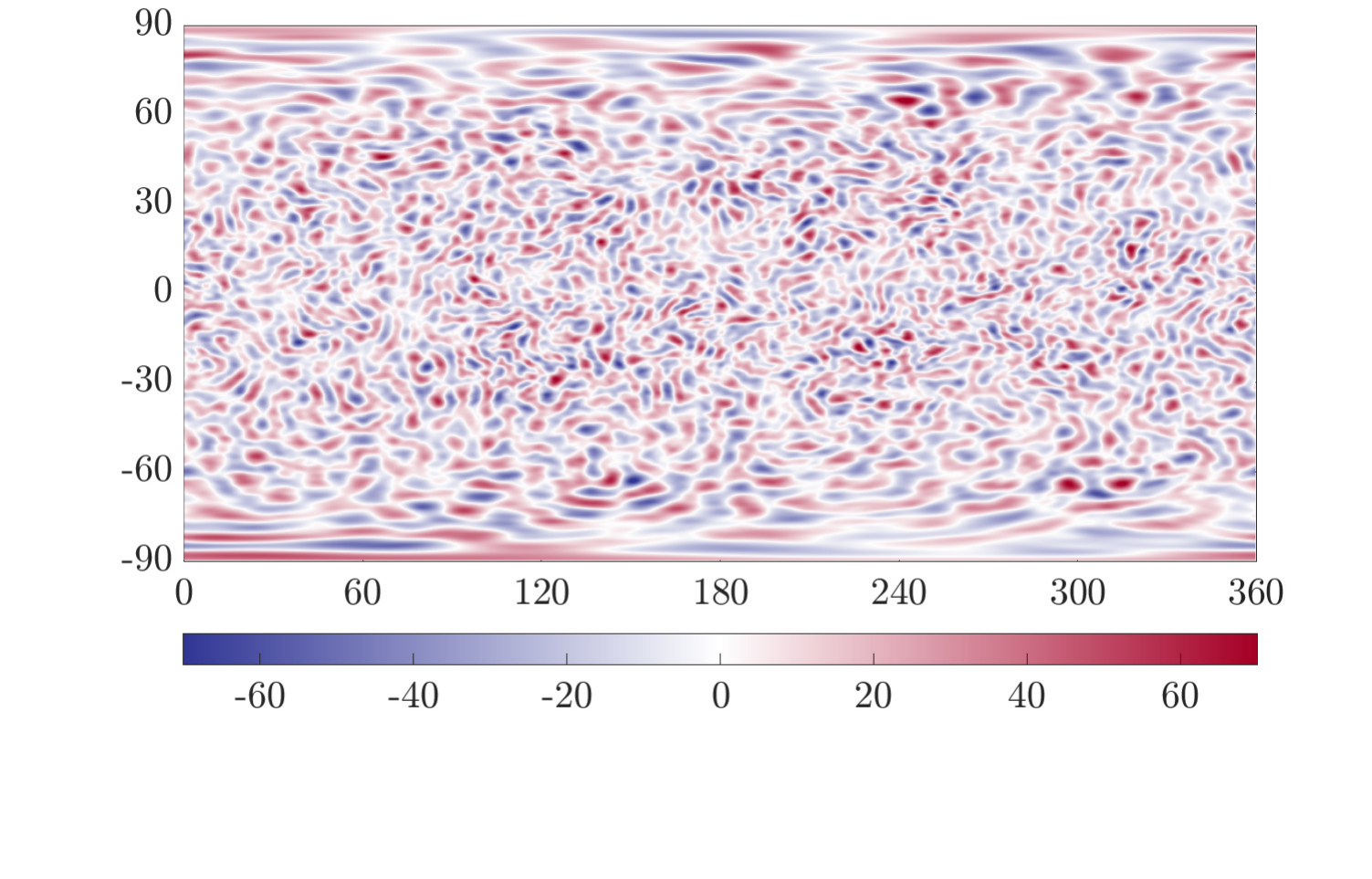}
    \end{subfigure}
    \begin{subfigure}{.47\textwidth}\centering
        \includegraphics[width=\columnwidth]{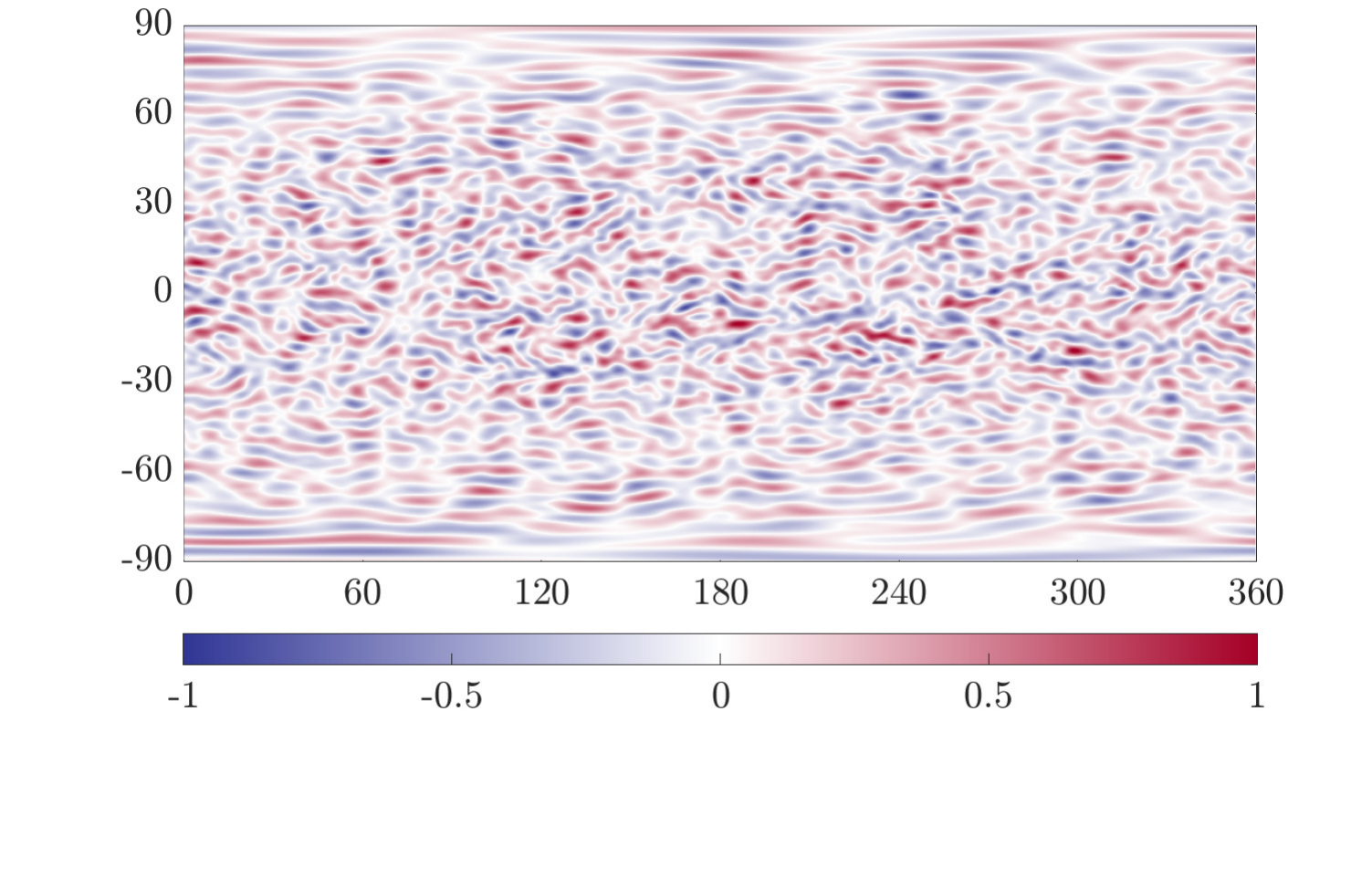}
    \end{subfigure}
    \begin{subfigure}{.47\textwidth}\centering
        \includegraphics[width=\columnwidth]{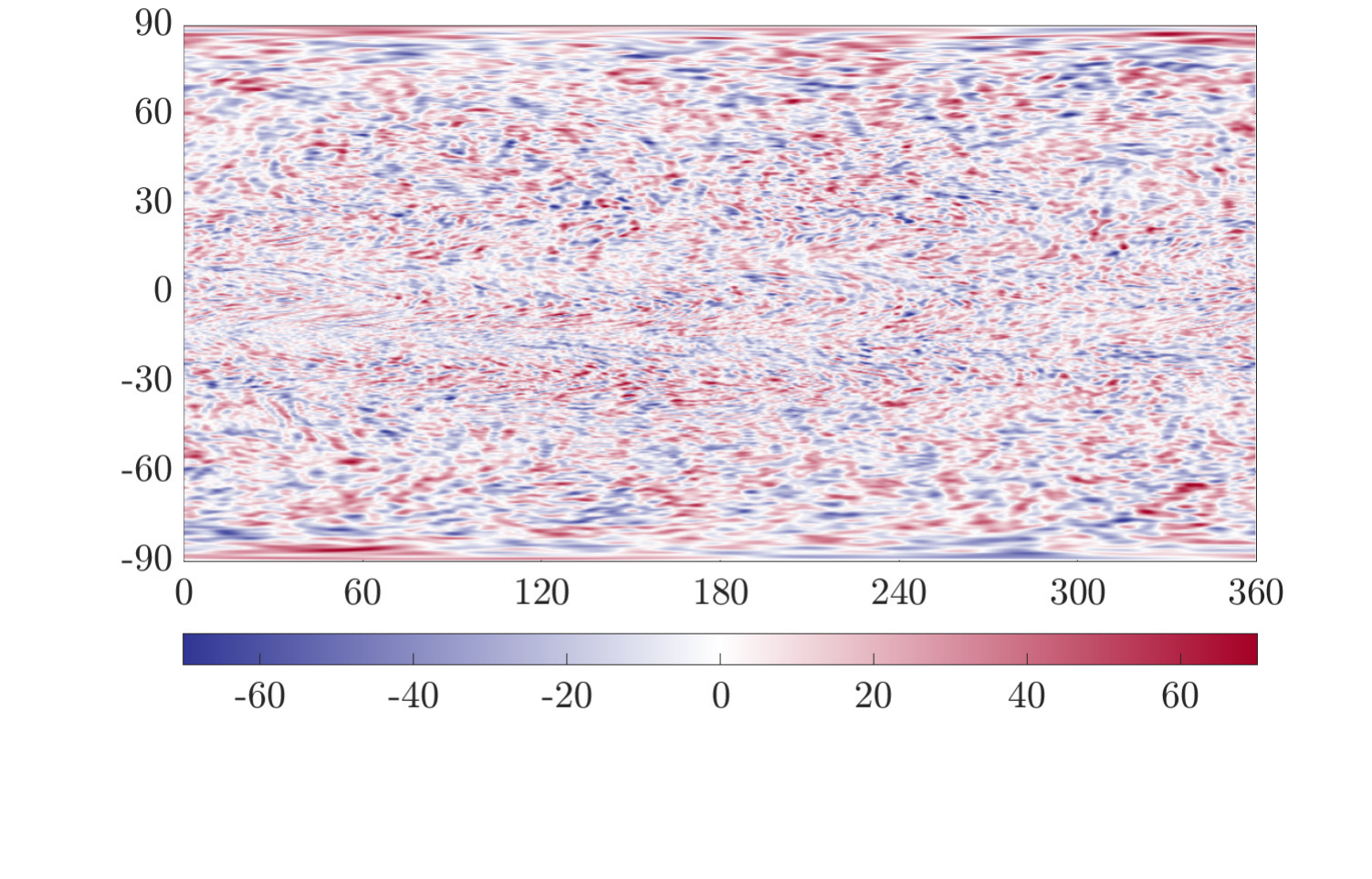}
    \end{subfigure}
    \begin{subfigure}{.47\textwidth}\centering
        \includegraphics[width=\columnwidth]{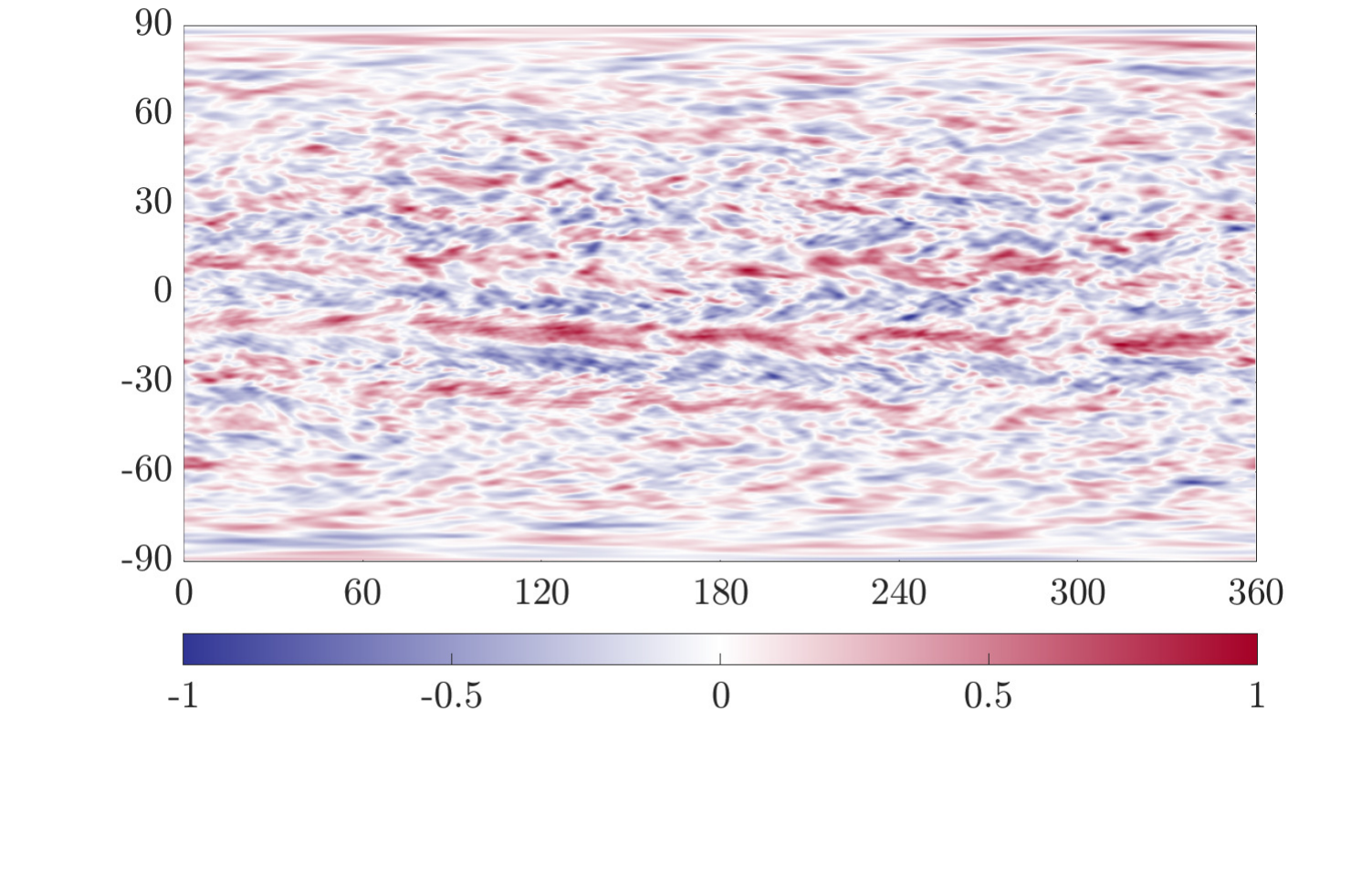}
    \end{subfigure}
    \begin{subfigure}{.47\textwidth}\centering
        \includegraphics[width=\columnwidth]{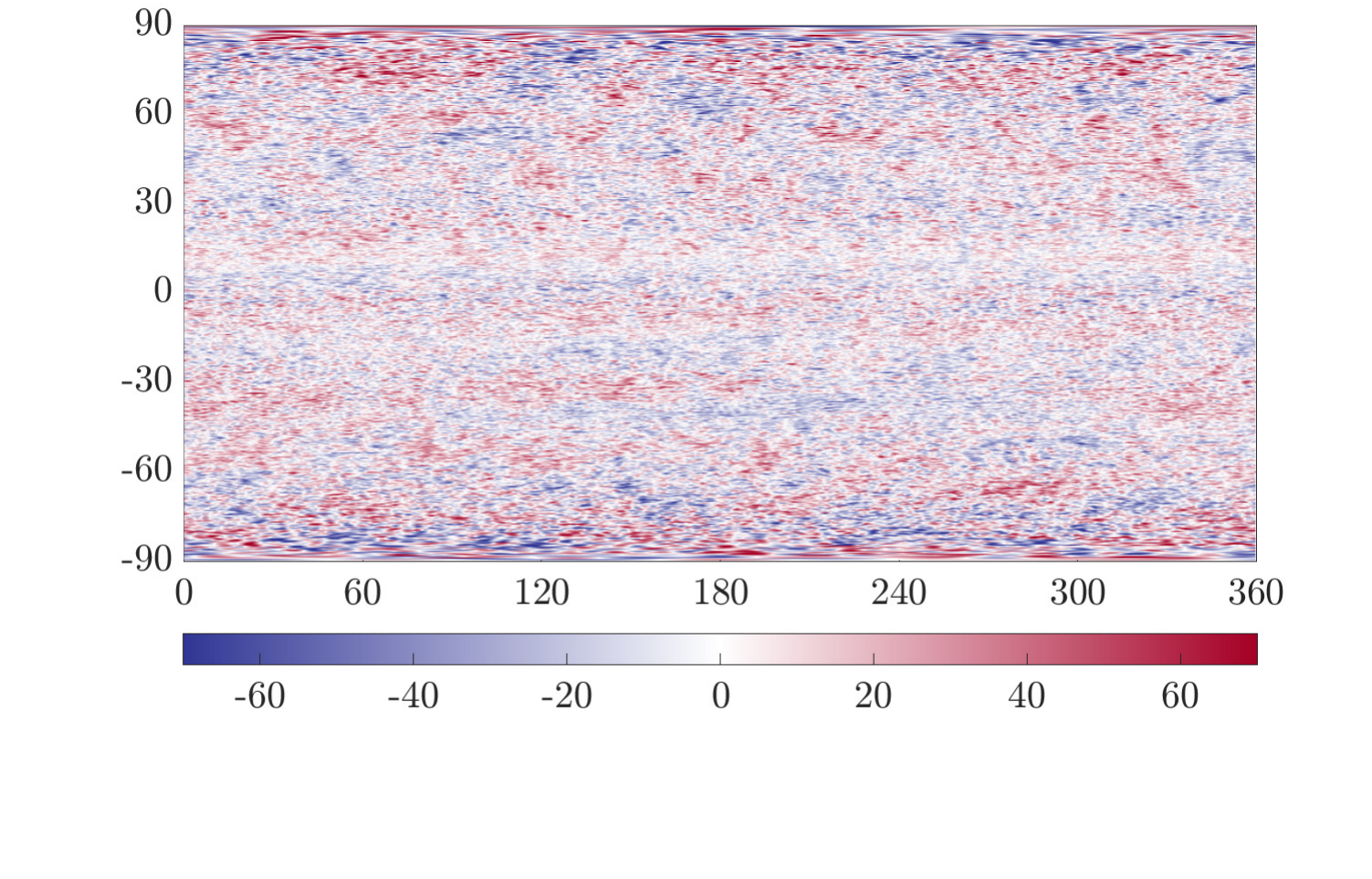}
    \end{subfigure}
    \begin{subfigure}{.47\textwidth}\centering
        \includegraphics[width=\columnwidth]{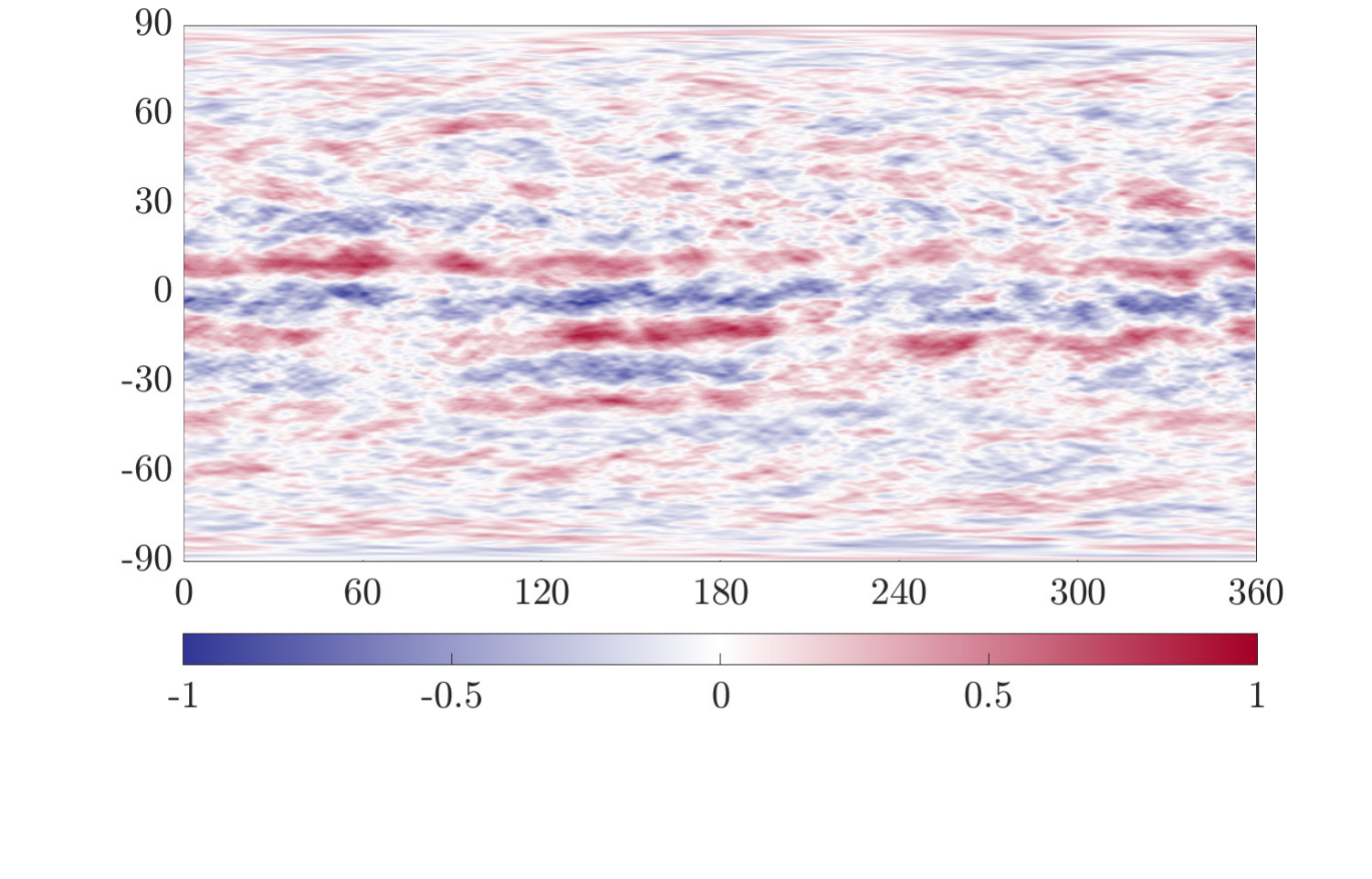}
    \end{subfigure}
    \begin{subfigure}{.47\textwidth}\centering
        \includegraphics[width=\columnwidth]{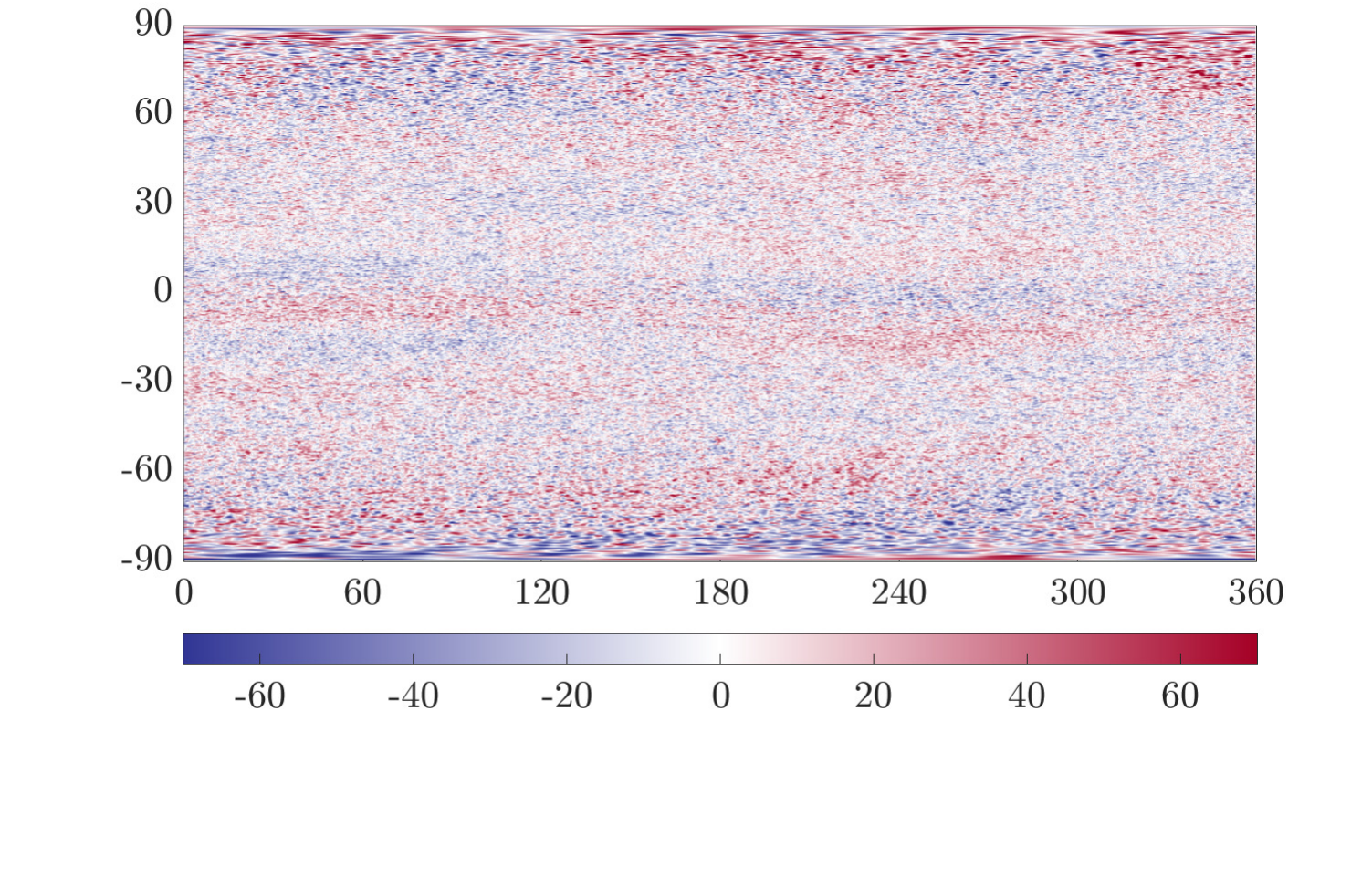}
    \end{subfigure}
    \begin{subfigure}{.47\textwidth}\centering
        \includegraphics[width=\columnwidth]{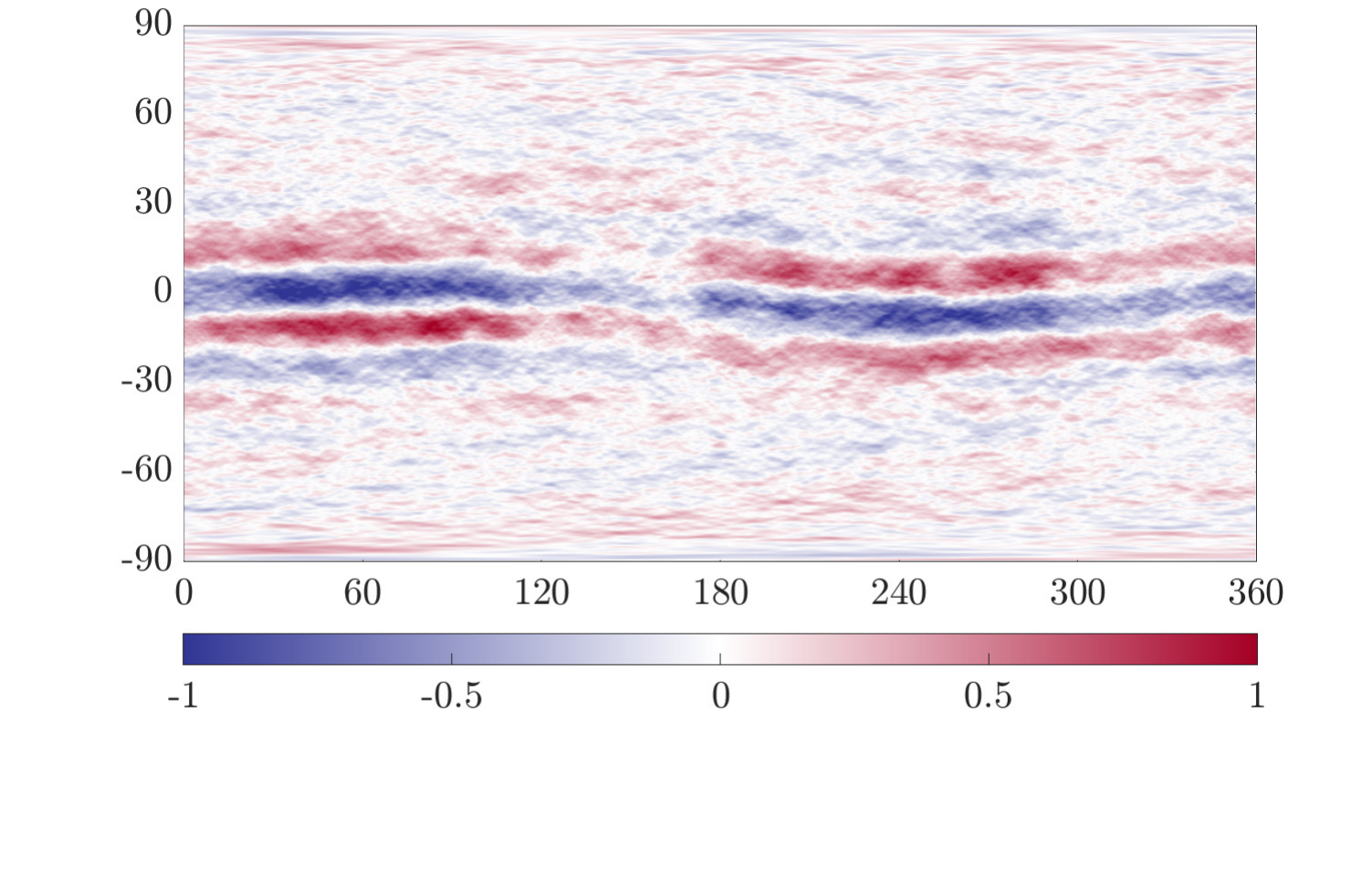}
    \end{subfigure}
    \caption{A sequence of solutions of freely evolving quasi-geostrophic flow on the sphere. Shown are the instantaneous potential vorticity anomaly (left column) and zonal component of the horizontal velocity field (right column) at the initial time (top row), after 100 days (second row), after 400 days (third row) and after 2000 days (bottom row).}
    \label{fig:snapshots}
\end{figure}

The development of these zonal jets is seen more clearly in Figure~\ref{fig:snapshots}, which shows the potential vorticity anomaly and the zonal velocity projected on a latitude-longitude grid. From top to bottom, it shows the development of the flow from the random initial condition to the formation of coherent structures in the longitudinal direction. After approximately 100 days, these structures form jets that circumnavigate the sphere, after which the flow settles into a steady flow pattern with three main jets in the equatorial region that contain the majority of the kinetic energy.

% \subsection{Snapshots of the flow}

These long-time simulations of turbulence in the absence of additional forcing are enabled by the fact that the discrete system shares the Lie-Poisson structure with the continuous model. The use of the symplectic isospectral time integrator ensures that the discrete Hamiltonian is approximately conserved, while all resolved Casimirs are preserved up to machine accuracy. To verify these properties, we define the relative error in the Hamiltonian $\widetilde{H}(t)$ and in the Casimirs $\widetilde{C}_n(t)$ as follows:

\begin{equation}
    \widetilde{H}(t) = \frac{H(t)-H(0)}{H(0)}, \quad \mbox{and} \quad \widetilde{C}_n(t) = \frac{C_n(t)-C_n(0)}{C_n(0)}, \quad n\geq 2.
\end{equation}

\begin{figure}
    \centering
    \begin{subfigure}{.47\textwidth}\centering
        \includegraphics[width=.8\columnwidth]{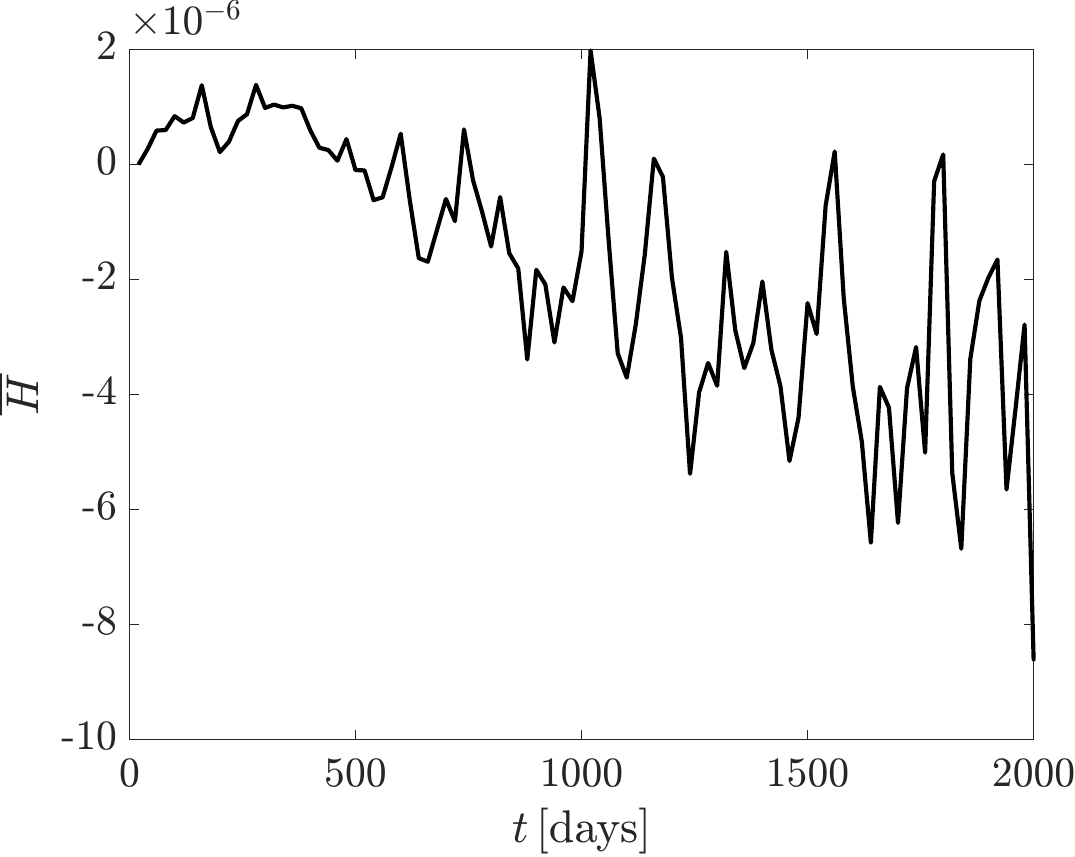}
    \end{subfigure}
    \begin{subfigure}{.47\textwidth}\centering
        \includegraphics[width=.8\columnwidth]{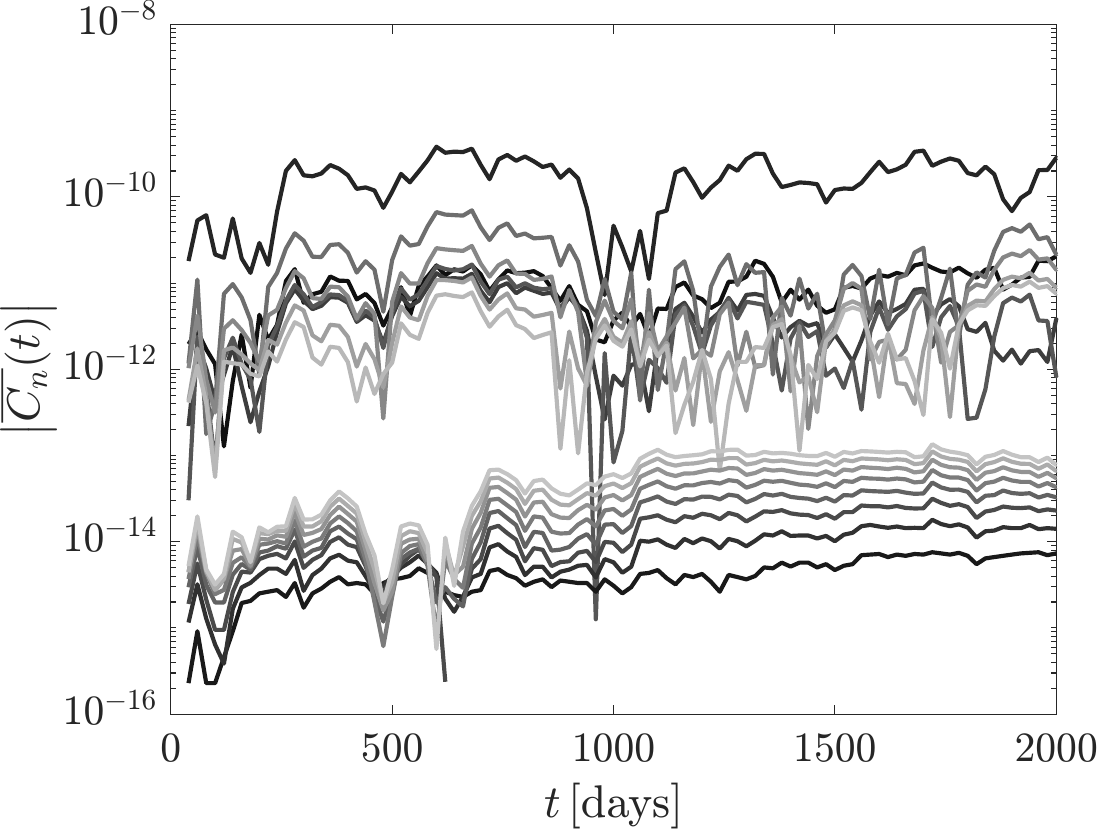}
    \end{subfigure}
    \caption{Time evolution in the relative deviation of the conserved quantities of the discrete system. On the left, the relative deviation of the discrete Hamiltonian from its initial value is shown. The right figure shows the relative errors in the first 16 Casimirs compared to their respective initial values. The different Casimirs are shown at a lighter color as the index increases. Notably, the Casimirs appear clustered according to the parity of their indices, with even indices being in the range $10^{-16}$-$10^{-13}$, and the odd indices in the range $10^{-12}$-$10^{-10}$.}
    \label{fig:Conservation}
\end{figure}

Figure~\ref{fig:Conservation} shows the time series for both the error in the Hamiltonian and the monomial Casimirs up to order $n=16$. It is well-known that the Hamiltonian cannot be generally exactly conserved, see \cite{zhong1988lie}. The Hamiltonian is conserved approximately at a relative error of $10^{-5}$. Casimirs of even parity are conserved up to machine accuracy, even at high monomial orders. There is a slight upward trend in the error as the monomial order becomes high. Round-off errors may be responsible for this deviation. More importantly, the long-time behaviour does not display any systematic trend. The Casimirs of odd parity are conserved up to an accuracy of $\mathcal{O}(10^{-10} - 10^{-12})$. The integrals involved in evaluating the odd-order Casimirs appear to be more senstive to round-off error in their evaluation, compared to the even-order Casimirs. As the monomial order increases, the relative errors show an approach toward values around $\mathcal{O}(10^{-13})$ for both odd and even orders.

\section{Conclusion and outlook}\label{sec:conclusion}
In this paper, we derived the spherical quasi-geostrophic equations from the rotating shallow water equations on the sphere via perturbation series methods in vorticity and divergence variables. The vorticity and divergence formulation of the rotating shallow water equations was obtained by means of the Hodge decomposition. Our derivation using methods of asymptotic analysis leads to the same quasi-geostrophic model as the works of \cite{verkley2009balanced, schubert2009shallow} with the added benefit of quantifying the conditions for the validity of the derivation. The result of the derivation is a global quasi-geostrophic potential vorticity model on the sphere without approximation of the Coriolis parameter, setting it apart from the well known $f$-plane and $\beta$-plane quasi-geostrophic models. The evolution equation for potential vorticity in combination with the relation between potential vorticity and stream function form a closed model. We showed that this model has a rich mathematical structure by deriving a Lagrangian description and a Hamiltonian description. The Hamiltonian description used the Lie-Poisson bracket of two-dimensional incompressible fluid dynamics, which enabled the characterisation of the geometric invariants of the quasi-geostrophic model.

The geometric properties of the quasi-geostrophic model motivate the use of recently developed structure-preserving methods to solve the model equations. We showed that the structure-preserving integrators are able to preserve Casimir invariants up to machine precision, which is important for capturing nonlinear interactions in the flow. The variational description and the Hamiltonian formulation also enable the development of variational integrators or energy-preserving integrators different from those used here, following approaches similar to \cite{brecht2019variational, wimmer2020energy}. One may formulate other structure-preserving time integrators. Current research is devoted to determining the evolution of the conservation error in the Hamiltonian as obtained by different variational time integrators to further optimise the long-term capturing of the flow dynamics. Another important benefit of having a geometric description of the model is the possibility of including stochastic transport parametrisation schemes of the type introduced in \cite{holm2015variational}. This enables the incorporation of data into the quasi-geostrophic model, which is of possible interest for simulations of Jovian atmospheres.

The analysis provided in this paper may serve as a point of departure for the derivation of geophysical fluid dynamical models with additional physical effects of which the thermal quasi-geostrophic equations are an important example. Further research is dedicated to systems in thermal geostrophic balance on the sphere, which have been analysed on flat domains recently in \cite{holm2021stochastic, crisan2023theoretical, beron2021nonlinear}.

\section*{Acknowledgements}
The authors gratefully acknowledge Francisco-Javier Beron-Vera (RSMAS Miami), Darryl Holm (Imperial College London), Klas Modin (Chalmers University of Technology and University of Gothenburg), Milo Viviani (SNS Pisa) and Paolo Cifani (SNS Pisa) for the many fruitful and inspiring discussions. The simulations were made possible through the Multiscale Modeling and Simulation computing grant of the Dutch Science Foundation (NWO) and carried out on the Dutch national e-infrastructure with the support of SURF Cooperative. EL was supported by NWO grant VI.Vidi.213.070.

\section*{Appendix: spherical geometry}\label{sec:spherical}
In this appendix, we introduce the coordinate system and the vector calculus expressions in spherical coordinates that we use in the derivations in Section \ref{sec:derivation}. Consider $\mathbb{R}^3$ with spherical coordinates $(\lambda,\theta,r)$ , where $\lambda \in [0,2\pi)$ is the angular distance eastwards (i.e., longitude), $\theta \in [-\pi/2, \pi/2]$ is the angular distance polewards (i.e., latitude) measured from the equator and $r$ is the radial distance from the origin in $\mathbb{R}^3$, which is taken as the center of the sphere. The common expression for the line element in the context of geophysical fluid dynamics, see for instance \cite{vallis2017atmospheric}, is
\begin{equation}\label{eq:ds1}
    ds^2 = dr^2 + r^2d\theta^2 + r^2\cos^2\theta \,d\lambda^2.
\end{equation}
In quantum mechanics, acoustics and numerous other applications, the line element is given by
\begin{equation}\label{eq:ds2}
    ds^2 = dr^2 + r^2d\theta^2 + r^2\sin^2\theta\,d\lambda^2.
\end{equation}
This difference arises because in geophysical fluid dynamics, the Equator is identified with zero latitude, whereas in other application areas, the North pole would be at zero latitude. Although (\ref{eq:ds1}) and (\ref{eq:ds2}) are fully equivalent, we select the spherical coordinate system \eqref{eq:ds2} for convenience, to comply with the standard use of spherical harmonics that are instrumental to the numerical integration of the spherical quasi-geostrophic equations. This coordinate system is known as the colatitude system with $\theta\in[0,\pi]$ and the equator located at $\theta = \frac{\pi}{2}$. The sphere is a two-dimensional smooth manifold $S^2\subset\mathbb{R}^3$ that can be obtained by considering level sets $r = a>0$ of the radial coordinate. The area form is then given by
\begin{equation}
    dA = a^2\sin\theta \,d\theta\,d\lambda.
\end{equation}

In two-dimensional calculus, the operations one encounters are the gradient, the perpendicular gradient, the divergence and the perpendicular divergence. In symbolic notation, two-dimensional calculus uses the set of operators $\{\nabla,\star,\cdot\}$, where $\nabla$ is the gradient operator, $\star$ is the Hodge star operator (more details can be found in \cite{flanders1963differential}), and $\cdot$ is the scalar product. The Hodge star operator depends on the metric and the orientation of the underlying manifold, but in the present context the Hodge star operator can be viewed as the operator $\perp$ that assigns to a vector its right-handed orthogonal, i.e., $(x_1,x_2)^\perp = (-x_2,x_1)$ and $\nabla^\perp = (\partial_1,\partial_2)^\perp = (-\partial_2,\partial_1)$. Together, the aforementioned operators form two dual de Rham complexes
\begin{center}
\begin{tikzcd}
C^\infty(S^2) \arrow[r,"\nabla"] \arrow[d,"\star"] & C^\infty(S^2,\mathbb{R}^2) \arrow[r,"\nabla^\perp\cdot"] \arrow[d,"\star"] & C^\infty(S^2) \arrow[d,"\star"]\\
C^\infty(S^2) & C^\infty(S^2,\mathbb{R}^2)  \arrow[l,"\nabla\cdot"'] & C^\infty(S^2) \arrow[l,"\nabla^\perp"']
\end{tikzcd}
\end{center}

The de Rham complexes give an interpretation of various important vector calculus identities and play a central role in defining the Hodge decomposition. De Rham complexes are instrumental in the design of finite element exterior calculus (\cite{arnold2018finite}), compatible finite element methods (\cite{cotter2023compatible}), discrete exterior calculus (\cite{hirani2003discrete}) and other structure-preserving algorithms, which are promising alternatives to the isospectral Lie-Poisson method employed in this work. Structure-preserving algorithms are particularly important for geophysical fluid dynamics and climate where stable long-time simulations are necessary without pollution by numerical dissipation.

On a sphere of radius $a$ parametrised by colatitude $\theta$ and longitude $\lambda$, the operators in the de Rham complex take the following expressions in coordinates. Let $f(\theta,\lambda)\in C^\infty(S^2)$ and $u(\theta,\lambda)=u_\theta(\theta,\lambda)\hat{\theta} + u_\lambda(\theta,\lambda)\hat{\lambda}\in C^\infty(S^2,\mathbb{R}^2)$, where $\hat{\theta},\hat{\lambda}$ are the unit vectors in the colatitudonal direction and the longitudonal direction, respectively. Then
\begin{equation}
\begin{aligned}
    \nabla f(\theta,\lambda)&= \frac{1}{a}\frac{\partial f}{\partial \theta}\hat{\theta} + \frac{1}{a\sin\theta}\frac{\partial f}{\partial \lambda}\hat{\lambda},\\
    \nabla^\perp f(\theta,\lambda)&= -\frac{1}{a\sin\theta}\frac{\partial f}{\partial \lambda}\hat{\theta} + \frac{1}{a}\frac{\partial f}{\partial \theta}\hat{\lambda},\\
    \nabla\cdot u(\theta,\lambda)&= \frac{1}{a\sin\theta}\frac{\partial}{\partial \theta}\left(u_\theta \sin\theta \right) + \frac{1}{a\sin\theta}\frac{\partial u_\lambda}{\partial \lambda},\\
    \nabla^\perp\cdot u(\theta,\lambda)&= -\frac{1}{a \sin\theta}\frac{\partial u_\theta}{\partial \lambda} + \frac{1}{a\sin\theta}\frac{\partial}{\partial \theta}(u_\lambda\sin\theta).
\end{aligned}
\end{equation}
In particular, it can be shown by direct calculation that $\nabla^\perp\cdot\nabla = 0$ and $\nabla\cdot\nabla^\perp=0$. In the derivation of the above identities, it is important to note that on the sphere the metric is not constant and so the area form needs to be taken into account whenever integration by parts is performed. These identities are two dimensional analogues of the well-known three-dimensional calculus identities $\nabla\cdot\nabla\times = 0$ and $\nabla\times\nabla = 0$. The scalar and vector Laplacian on the sphere are 
\begin{equation}
    \begin{aligned}
            \Delta f(\theta,\lambda) &= \nabla\cdot\big(\nabla f(\theta,\lambda)\big) = \nabla^\perp\cdot\big(\nabla^\perp f(\theta,\lambda)\big) \\
            &= \frac{1}{a^2\sin\theta} \frac{\partial}{\partial \theta}\left(\sin\theta\frac{\partial f}{\partial \theta}\right) + \frac{1}{a^2 \sin^2\theta}\frac{\partial^2 f}{\partial \lambda^2},\\
            \Delta u(\theta,\lambda) &= \nabla\big(\nabla\cdot u(\theta,\lambda)\big) + \nabla^\perp\big(\nabla^\perp\cdot u(\theta,\lambda)\big),
    \end{aligned}
\end{equation}
where one can expand the definition of the vector Laplacian to obtain the usual unwieldy coordinate expression. The Hodge decomposition states that a vector field $u\in C^\infty(S^2,\mathbb{R}^2)$ can be decomposed as 
\begin{equation}\label{eq:hodge}
u = \nabla^\perp \psi + \nabla \chi,
\end{equation}
where $\psi,\chi\in C^\infty(S^2)$. The $\psi$ and $\chi$ are elements of, respectively, the bottom right corner and top left corner of the two de Rham complexes above. It follows from the Hodge decomposition that an incompressible vector field can be expressed completely in terms of a single function. In the context of fluid dynamics this function is called the stream function. The scalar transport equation in spherical coordinates for a function $f\in C^\infty(S^2,\mathbb{R}^2)$ is given by
\begin{equation}\label{eq:transport}
    \frac{D}{Dt}f = \frac{\partial f}{\partial t} + u\cdot\nabla f = 0,
\end{equation}
where $\frac{D}{Dt}$ denotes the material derivative. Under the condition that $u$ is divergence-free, the transport equation can be expressed as
\begin{equation}\label{eq:transportandpb}
\begin{aligned}
    0 &= \frac{\partial f}{\partial t} + \nabla^\perp \psi\cdot\nabla f\\
    &= \frac{\partial f}{\partial t} + \{\psi,f\}\\
    &= \frac{\partial f}{\partial t} + \frac{1}{a^2\sin\theta}\left(\frac{\partial \psi}{\partial \theta}\frac{\partial f}{\partial \lambda}-\frac{\partial f}{\partial \theta}\frac{\partial \psi}{\partial \lambda}\right).
\end{aligned}
\end{equation}
In \eqref{eq:transportandpb}, one finds the relations between transport and the Poisson bracket $\{\,\cdot\,,\,\cdot\,\}:C^\infty(S^2)\times C^\infty(S^2)\to C^\infty(S^2)$, and the coordinate expression of transport on the sphere. These identities play a fundamental role in fluid dynamics and the numerical discretisation of transport problems. We introduce one more relation that is convenient for transforming the equations of fluid dynamics from the advective form to the vector invariant form. This relation is called the fundamental vector identity of fluid dynamics in \cite{crisan2017mathematics} and is given by
\begin{equation}\label{eq:fvifd}
    u\cdot\nabla v + (\nabla u)^T\cdot v = (\nabla^\perp\cdot u)v^\perp + \nabla(u\cdot v).
\end{equation}
This identity is obtained by equating the two definitions of the Lie derivative operator in differential topology, see \cite{holm2021stochastic} for a discussion. In the next section we make frequent use of the relations introduced above in our derivation of the quasi-geostrophic potential vorticity on the sphere.

\bibliographystyle{plainnat}
\bibliography{biblio}

\end{document}